\documentclass[]{emulateapj-rtx4}

\usepackage{times}
\usepackage{epsfig}

\usepackage{longtable}
\usepackage{rotating}
\usepackage[usenames, dvipsnames]{color}   %Options added for more colors & names

\DeclareMathAlphabet{\mathsc}{OT1}{cmr}{m}{sc}
\def\testbx{bx}%
\DeclareRobustCommand{\ion}[2]{%
\relax\ifmmode
\ifx\testbx\f@series
{\mathbf{#1\,\mathsc{#2}}}\else
{\mathrm{#1\,\mathsc{#2}}}\fi
\else\textup{#1\,{\mdseries\textsc{#2}}}%
\fi}

\newcommand{\rsun}{\mbox{\,$\rm R_{\odot}$}}        % solar radius

\shorttitle{The outer corona}
\shortauthors{Del Zanna et al.}

\begin{document}

\title{Predicting the COSIE-C Signal from the Outer  Corona up to 3 Solar Radii. }

\author{Giulio Del Zanna}
\affil{DAMTP, CMS, University of Cambridge, Wilberforce Road, Cambridge CB3 0WA, United Kingdom}
 \email{gd232@cam.ac.uk}

%\author{Ed DeLuca}

\author{John Raymond}
\affil{Harvard-Smithsonian Center for Astrophysics, 60 Garden Street, Cambridge, MA 02138, United States}

\author{Vincenzo Andretta}
\affil{INAF/Osservatorio Astronomico di Capodimonte, Salita Moiariello 16, 80131 Napoli, Italy}

\author{Daniele Telloni}
\affil{INAF/Osservatorio Astrofisico di Torino, Strada Osservatorio 20, 10025 Pino Torinese (TO), Italy}

\author{Leon Golub}
\affil{Harvard-Smithsonian Center for Astrophysics, 60 Garden Street, Cambridge, MA 02138, United States}

\begin{abstract}
{We present estimates of the signal to be expected in quiescent solar conditions,
as would be obtained with the COronal Spectrographic Imager in the EUV 
 in its coronagraphic mode (COSIE-C).
COSIE-C has been proposed to routinely observe  the relatively 
unexplored outer corona, where we know that  many fundamental processes affecting 
both the lower corona and the solar wind are taking place.
The COSIE-C spectral band, 186--205~\AA, is well-known as it has been observed with  
Hinode EIS. We present Hinode EIS observations that we obtained in 
2007 out to  1.5~\rsun, 
to show that this spectral band in  quiescent streamers  is dominated by  \ion{Fe}{xii} and  \ion{Fe}{xi}
and that the  ionization temperature is nearly constant. 
 To estimate the COSIE-C signal in the  1.5--3.1~\rsun\ region we use a model based on  CHIANTI atomic data and 
SoHO UVCS observations in the  \ion{Si}{xii} and \ion{Mg}{x} coronal lines  of two quiescent 1996 streamers.
 We reproduce the observed EUV radiances with a simple density model,
photospheric abundances, and a constant temperature of 1.4 MK. We show that other theoretical 
or semi-empirical models fail to reproduce the observations.  
We find that the coronal COSIE-C  signal at  3~\rsun\ should be 
about 5 counts/s  per 3.1\arcsec\ pixel in quiescent streamers.
This is unprecedented and opens up a significant discovery space.
We also briefly discuss stray light  and the visibility of other solar features.
In particular, we present  UVCS observations of an active region 
streamer, indicating increased signal compared to the quiet Sun cases. 
}
\end{abstract}


%________________________________________________________________
\section{Introduction}

It is now quite well established that the outer corona, i.e. the region 
between 1.5 and 3~\rsun\ (as measured from Sun center)
is the place  where many fundamental processes are taking place. 
For example, this is the region where the solar wind and the 
Coronal Mass Ejections (CME) become accelerated, see e.g.
\cite{abbo_etal:2016, zhang_dere:2006, temmer_etal:2017}.
  This is also the region where CME driven
waves steepen into shocks, and as the Alfven Mach number increases from 1, the
post shock plasma in a CME can transition from low beta to high beta.

The outer corona is also the region where the small-scale complex topology of the 
magnetic field, which shapes the observed features of the lower corona,
becomes more simple and radial. 
It is also  a transition region from collisional fluid to collisionless plasma.

The outer corona could also contain, depending on which  model one considers, the 
transition region from the low-$\beta$ collisional inner corona
to the high-$\beta$ collisionless heliosphere and the nascent solar wind.
For example, Fig.~6 of the MHD model of \cite{vasquez_etal:2003}
shows that  $\beta=1$ around 2\rsun\ in a streamer,
while $\beta$ in a coronal hole is very small out to the edge of the plot at 4~\rsun. 
Other models put the  $\beta=1$ further out, closer to the Alfven point.

The outer corona is also the region where processes 
such interchange reconnection between closed and open structures is likely
taking place, even in the quiet Sun \citep[see, e.g.][]{fisk:2003}.
The complex mix of  quiet Sun areas and coronal holes 
creates a complex topological system.  MHD simulations of the 
quiet corona around the time of the 2008 eclipse run by the 
Predictive Science group with the MAS code \cite{mikic_etal:2007}
showed the presence of 
a multitude of separators and quasi-separatrix layers in the outer
corona, which have a profound effect on the magnetic connectivity 
between a point in the heliosphere and its source region,
the so-called S-web \citep{antiochos_etal:2011}.

When even a single active region is present, the topology of the global magnetic
field of the outer corona  becomes more complex.
For example, magnetic field modeling of a few isolated active regions observed in 2007
showed the presence of null points at about 2~\rsun\ 
\citep{delzanna_etal:2011_outflows}. A model of interchange reconnection 
taking place at these null points was able to reproduce 
 the location and strength \citep[see also][]{bradshaw_etal:11}
of the so-called `coronal outflows', regions mostly connected to sunspots 
that show upflowing plasma at temperatures above 2~MK 
\citep[see, e.g.][]{delzanna:08_flows,harra_etal:08,doschek_etal:08}. 
This  interchange reconnection process in active regions is in principle capable
of injecting coronal plasma that was originally present in the closed
hot (3 MK)  loops of active region cores into the heliosphere, and 
forming as a by-product the large-scale cooler (1 MK) loops which fan out of
sunspots and connect to the entire surroundings of an active region. 

It is  clear that the density at 2~\rsun\ is so low that 
whatever occurs  at  2~\rsun\ is virtually invisible in on-disk observations,
against the bright background of the inner corona.
Therefore, off-limb high-resolution observations around 2~\rsun\ are needed.
Yet, this region is relatively unexplored, with the exception of a 
few ground-based observations during total eclipses, and SoHO/UVCS \citep{uvcs}
spectra. Such observations returned plenty of scientific 
results, but were not capable of monitoring the  dynamical evolution of the outer corona.

Narrow-band high-resolution images of the solar corona in the visible taken during total eclipses
have  shown us the outer corona in its full glory, with many 
complex open and closed structures, and features that are difficult to comprehend.
For example, intriguing features have appeared in recent images of the 
\ion{Fe}{xi} visible forbidden line \citep{habbal_etal:2011}.
Such images of the coronal lines show more clearly the outer structures
not only because of the natural occulter (the Moon), but also because 
the intensities  of the visible forbidden lines decay more slowly 
 with height, compared to the allowed lines.
This occurs  partly because  forbidden line intensities  decay linearly with  the electron density,
and partly because the photospheric radiation increases their intensities
via photo-pumping \citep[see, e.g.][ for a recent discussion of visible and infrared 
forbidden lines]{delzanna_deluca:2017}.
Unfortunately, these eclipse images are very short snapshots of a corona that 
we know is highly  dynamic.

Above  2~\rsun,  coronagraphic observations  in the visible with e.g. the
 SoHO Large Angle Spectroscopic COronagraph (LASCO) C2 
\citep{brueckner_etal:1995}, and further out from  STEREO HI,  have provided 
valuable information about the dynamic outer corona. However, the 
spatial resolution and sensitivity have not been comparable with 
ground-based eclipse observations. Future coronagraphs in the visible
(e.g. Proba-3, Aditya) will provide significant improvements,
but such instruments will always be limited by the fact
that  the coronal signal is  more than 6 orders of magnitude weaker 
than the photospheric one. 
The Solar Orbiter Metis coronagraph will be an improvement on the current 
situation as it will 
provide images both in the visible and in the UV, in the hydrogen 
Ly $\alpha$ above 1.6~\rsun.  The Solar Orbiter EUI full-disk images are in principle also 
capable of observing the outer corona, but outside the currently planned 
remote-sensing windows (during close encounter).

Around 2~\rsun, we had only a few  images of the outer corona. 
LASCO/C1 observed the outer corona up to 3\rsun\ in the 
green and red visible forbidden lines, but returned only limited
results \cite[see, e.g.][]{mierla_etal:2008}.
We have plenty of observations from the X-rays to the UV of the lower corona
in a range of spectral lines and broad-bands, 
typically up to 1.3~\rsun, 
but the nearly exponential decay of the electron density
with radial distance, typical of stationary isothermal plasma,
 reduces dramatically the XUV signal, as the 
intensities of dipole-allowed transitions are proportional  
to the square of the electron density.

There are a few off-points from SDO AIA (mainly for comet
observations) where we see  structures in the EUV  out beyond 2\rsun.
Recently, also SUVI obtained some off-limb images. 
We also have PROBA2/SWAP EUV images (in a band around 174~\AA\ dominated by 
\ion{Fe}{x} and   \ion{Fe}{ix}) which clearly show a very complex and dynamical  behavior of 
open and closed structures, especially above active regions. 
 The dynamical behavior is mainly seen in the SWAP Carrington movies. 
The effects of an emerging AR are also clearly evident
in AIA, see e.g.  \cite{schrijver_higgins:2015}.

To overcome the lack of detailed information about the 
outer corona, an instrument of novel design has recently been proposed to fly on the 
International Space Station: the COronal Spectrographic Imager in the EUV (COSIE).  
The instrument has two modes of operation. The spectrograph mode will 
mainly be used  on-disk: it is a full Sun slitless
imaging spectrograph (COSIE-S). 
The other one is a broad-band imager which will mainly be used in its
coronagraphic mode (COSIE-C), although it can also observe on-disk, with a filter
to reduce the bright on-disk signal.

The key issue here is that building a coronagraph in the visible has always
been a challenge, given the huge dynamic range between the disk intensity and 
the signal of the outer corona. This is not the case in the EUV, 
so an instrument such as COSIE can naturally observe both on-disk 
and off-limb. 
 
In a nutshell,  COSIE-C  is designed to observe the corona in an EUV 
spectral band between 186 and 205~\AA\ with a high-cadence, a 
large field of view (FOV) of 6.6\rsun\ $\times$ 6.6\rsun, 
and a spatial resolution of 3.1\arcsec. 
Such a resolution is really high, when compared to previous coronagraph images, 
and comparable to that of many current EUV instruments observing on-disk, such as 
Hinode EIS.
This spectral range is well-known, as it has been routinely observed with 
many previous and current space-based instruments such as
SoHO/CDS, SoHO/EIT,  TRACE, STEREO/SECCHI, SDO/AIA, SDO/EVE,  GOES SUVI.
In particular, detailed spectroscopic observations with Hinode/EIS 
have been studied in detail, and over the past years all the main lines have been
identified, and the  atomic data 
has been recalculated and benchmarked, as reviewed in \cite{delzanna:12_atlas}.
The quiet solar corona (at about 1 MK) is emitting 
in this spectral region strong coronal lines from iron ions: 
\ion{Fe}{x}, \ion{Fe}{xi}, and \ion{Fe}{xii}.

The main aim of the coronagraphic mode of COSIE is to 
provide continuous monitoring of the outer corona with high-cadence and
sensitivity, so  dynamic events such as CME or waves in the
outer corona  can be studied in detail.

In order to achieve this, a key technical issue  regards the ability 
to observe simultaneously the inner corona from the solar limb to the
outer corona. 
In principle the instrument is capable of 
observing up to 4.6~\rsun\ from Sun center, along the diagonal of its square 
FOV of 6.6$\times$6.6~\rsun, however as we have discussed the most interesting
(and relatively unexplored) region of the solar corona is between 1.5 and 3.3~\rsun\, 
which is what will be observed within the square FOV of COSIE.

The main aim of the present paper is therefore to provide  observational 
 evidence of how the brightness of the EUV corona 
changes with radial distance from the Sun in the 1.5--3~\rsun\ range,
so we can provide relatively accurate estimates of the COSIE signal
up to these distances. 
We mainly consider quiet Sun  streamers in this paper as they are much less variable
than e.g. those above active regions, or signals in other features.
However, we provide one example of the signal above an active region streamer,
and provide some comments  of what signals could be expected for other
regions.

Our main aim might  seem relatively simple, from an observational or
modeling perspective, but it is not, as we discuss here. 
 We know, from past observations, that the intensity
of the disk in the EUV is comparable with the intensity of the inner
corona as observed off-limb, but the behavior of the coronal lines 
 up to about 3~\rsun\ was not known until the present study. 
Within the literature, we have not found  direct
measurements of coronal lines up to 3~\rsun\ in the quiet Sun.
The only study of the outer corona is 
that by \cite{goryaev_etal:2014}, where however 
a streamer above an active region was observed. 
PROBA2/SWAP images in the 174~\AA\ band  were analyzed up to 2~\rsun.
The broad-band signal decreased by about 3 orders
of magnitude in this range. Interestingly,  a nearly
isothermal corona of 1.4~MK was needed to explain the observations,
as we have also found here.

\begin{figure}[htbp]
 \centerline{\includegraphics[width=7.0cm, angle=0]{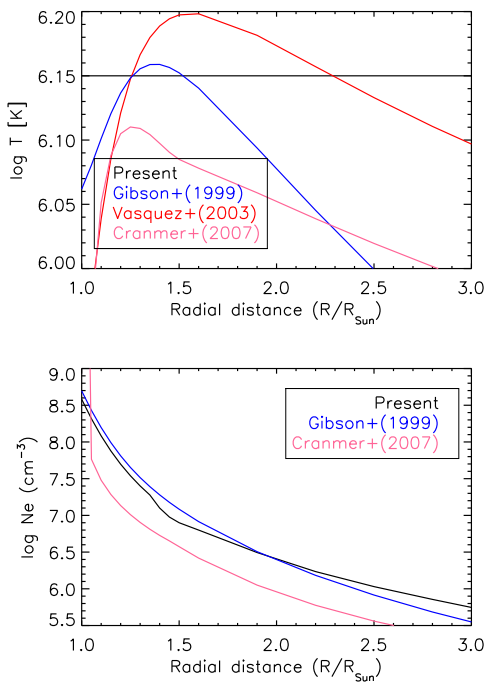}}
\caption{The  radial  profiles of the electron densities and temperatures 
we assumed for  modeling quiet Sun streamers (black lines), 
compared to some literature values.}
\label{fig:ne_te}
\end{figure}

In principle, we could supplement the lack of observations with 
modeling. However, being an unexplored region, we really do not 
know how the fundamental plasma parameters vary with radial 
distance. Even considering the most simple case of a 
streamer in the quiet Sun,  very different 
estimates of densities and temperatures have been published.
A few are shown in Figure~\ref{fig:ne_te}, together with those
we have adopted for our modeling, as described below.

% Figure~\ref{fig:ne_te}

Such variable  densities and temperatures 
 produce very different estimates of coronal line radiances, 
as discussed in \cite{andretta_etal:2012} and as also shown here. 
Furthermore, an issue which greatly complicates any 
modeling effort is the variability of the elemental 
abundances, as we discuss below.

An ideal instrument which explored the outer corona is   SoHO/UVCS 
 because it  observed many coronal lines formed at similar temperatures
as those that  contribute to the COSIE-C band.
There are, of course, many published results from  SoHO/UVCS (and 
the previous similar  SPARTAN instrument)
which observed the outer corona from 1996 even as far as 10~\rsun.
However, the instrument sensitivity was such that only the 
brightest lines, the H I Lyman $\alpha$ and the 
\ion{O}{vi} doublet,  had enough signal in the outer corona. 
 These lines are largely affected by photoexcitation as the density of the outer corona
decreases, so they are visible to greater distances,
but are not ideal to estimate the signal in the collisionally-dominated EUV lines,
as we will discuss below.

There are  many published results from UVCS, where 
several coronal EUV lines  were observed. 
However, they were mostly  at distances of  about 1.5~\rsun, see e.g.
\cite{raymond_etal:97,parenti_etal:2000}. 
We have therefore searched the UVCS database to try and find 
some observations that could be useful for our purpose.
We have identified a few observations
of the Mg X and Si XII coronal lines in the 1.4--3~\rsun\ range, which we could use to 
build a model to estimate the  COSIE Iron line
count rates  in the 186--205~\AA\ range. 
Several new and interesting results are obtained from this analysis.
At lower heights, we  present the analysis of a unique 
Hinode EIS off-limb 
observation where coronal lines (the same that would be observed by
COSIE) were visible up to  1.5~\rsun.

The paper is organized as follows: Section~2
describes the UVCS observations of quiescent streamers and their analysis.
Section~3 describes the way we have modeled 
the UVCS observed radiances, and the predictions we make for the 
COSIE-C signal in the outer corona.
 Section~4 briefly discusses the EIS off-limb 
observation,  while Section~5 briefly discusses stray light issue.
Section~6 provides some comments on the expected COSIE-C signal in
other regions and features, while Section~7  draws the conclusions.

\section{Quiet Sun UVCS  Observations and data analysis}

During 1996 and early 1997, the Sun was relatively quiet,
with the exception of a few small active regions, and a long-lived large one,
located in the southern hemisphere, at the end of a large, long-lasting
trans-equatorial coronal hole, the `Elephant's trunk'
\citep{delzanna_jgr99a}. 

During this period, the UVCS instrument was routinely running 
synoptic studies, where the corona between about 1.4 and 3.5\rsun
was observed by moving the slit in 4-5 different radial locations 
and then rotating the slit around the Sun. 
Coronagraph and UVCS observations during this period clearly show
how the streamers are affected by the presence of these different
structures, especially active regions. 
A review summary of the UVCS observations in the H I and the  O VI lines
during this period  can be found in \cite{antonucci_etal:2005}.

\begin{figure*}[!htbp]
 \centerline{
\includegraphics[width=5.0cm,angle=0]{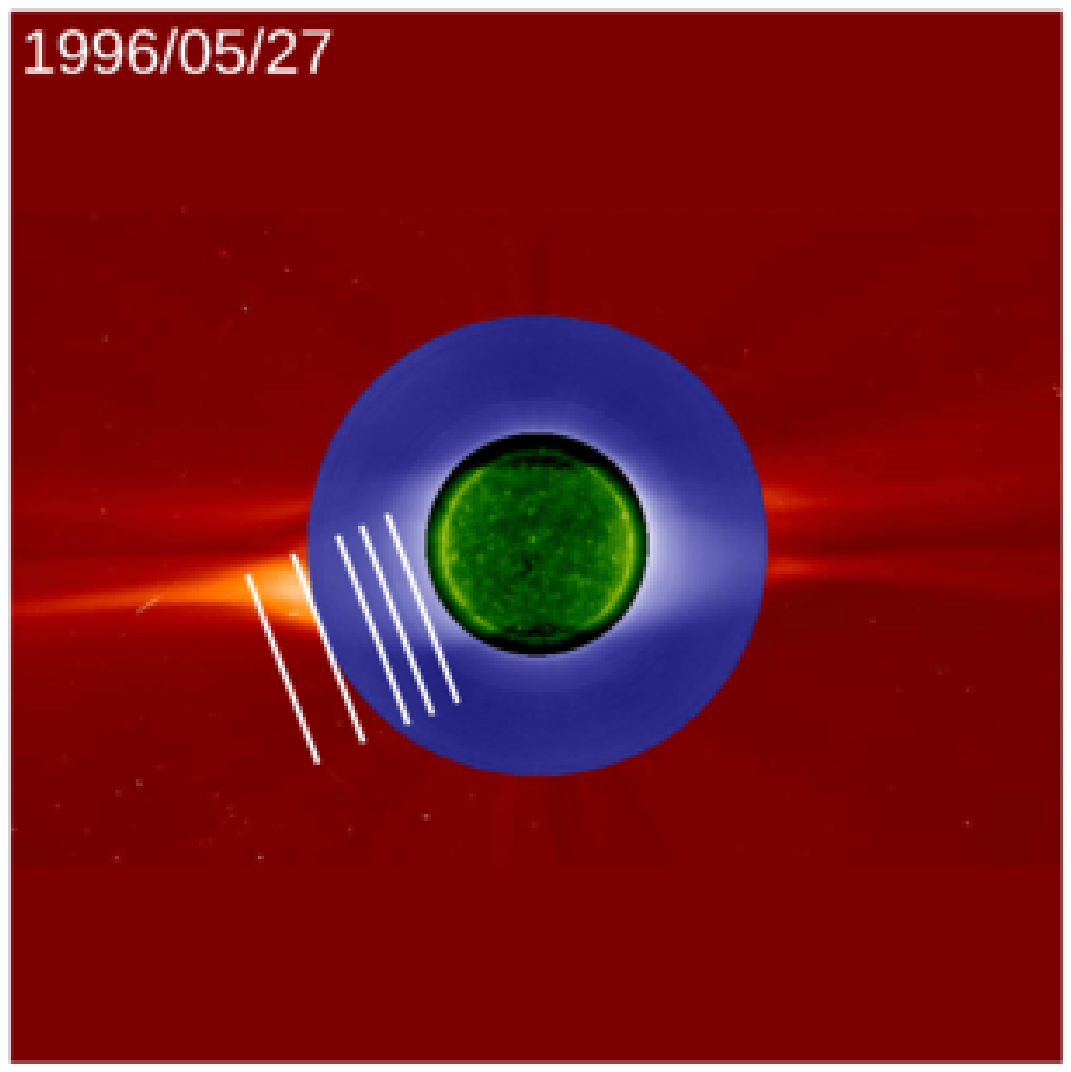}
\includegraphics[width=5.0cm,angle=0]{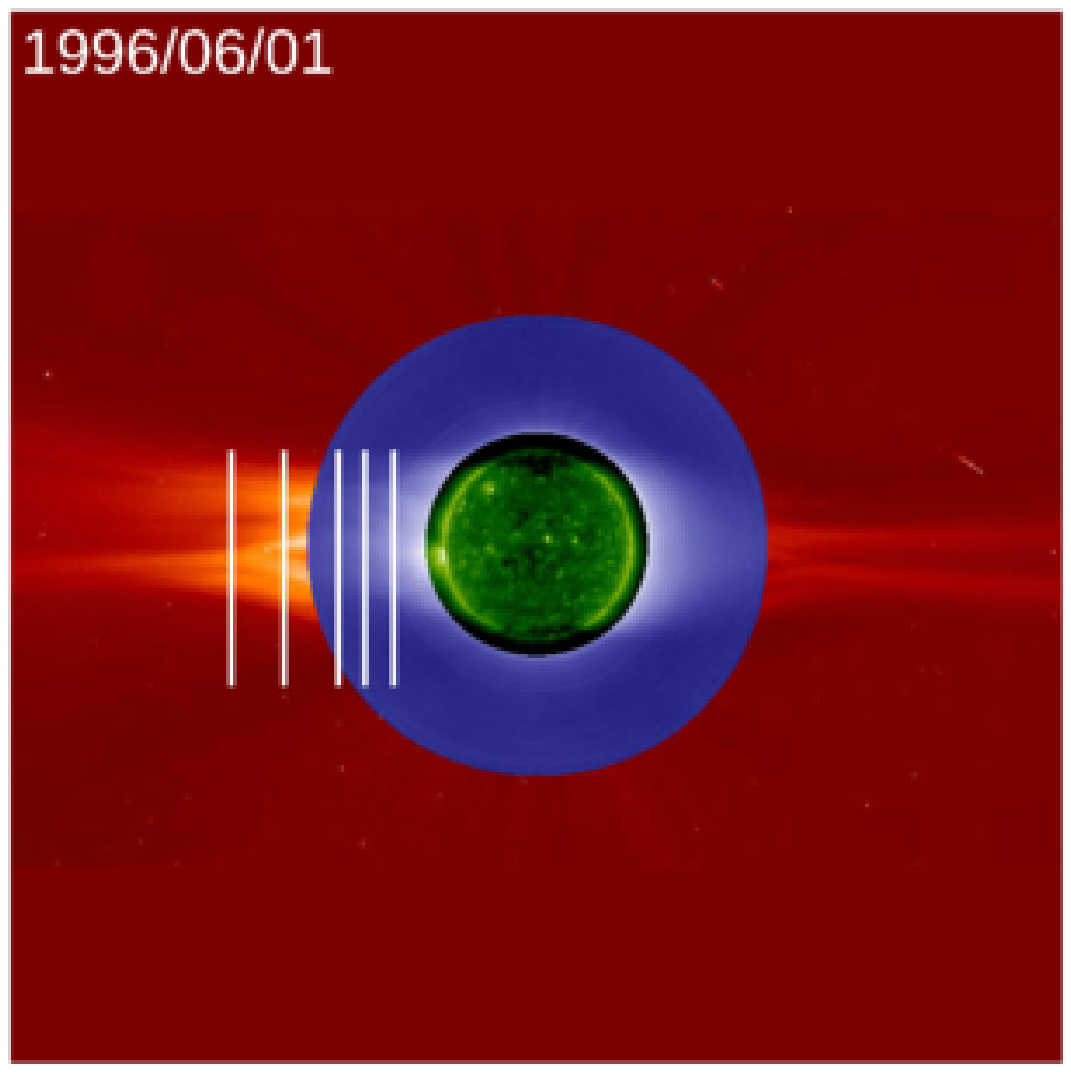}}
 \centerline{
\includegraphics[width=5.0cm,angle=0]{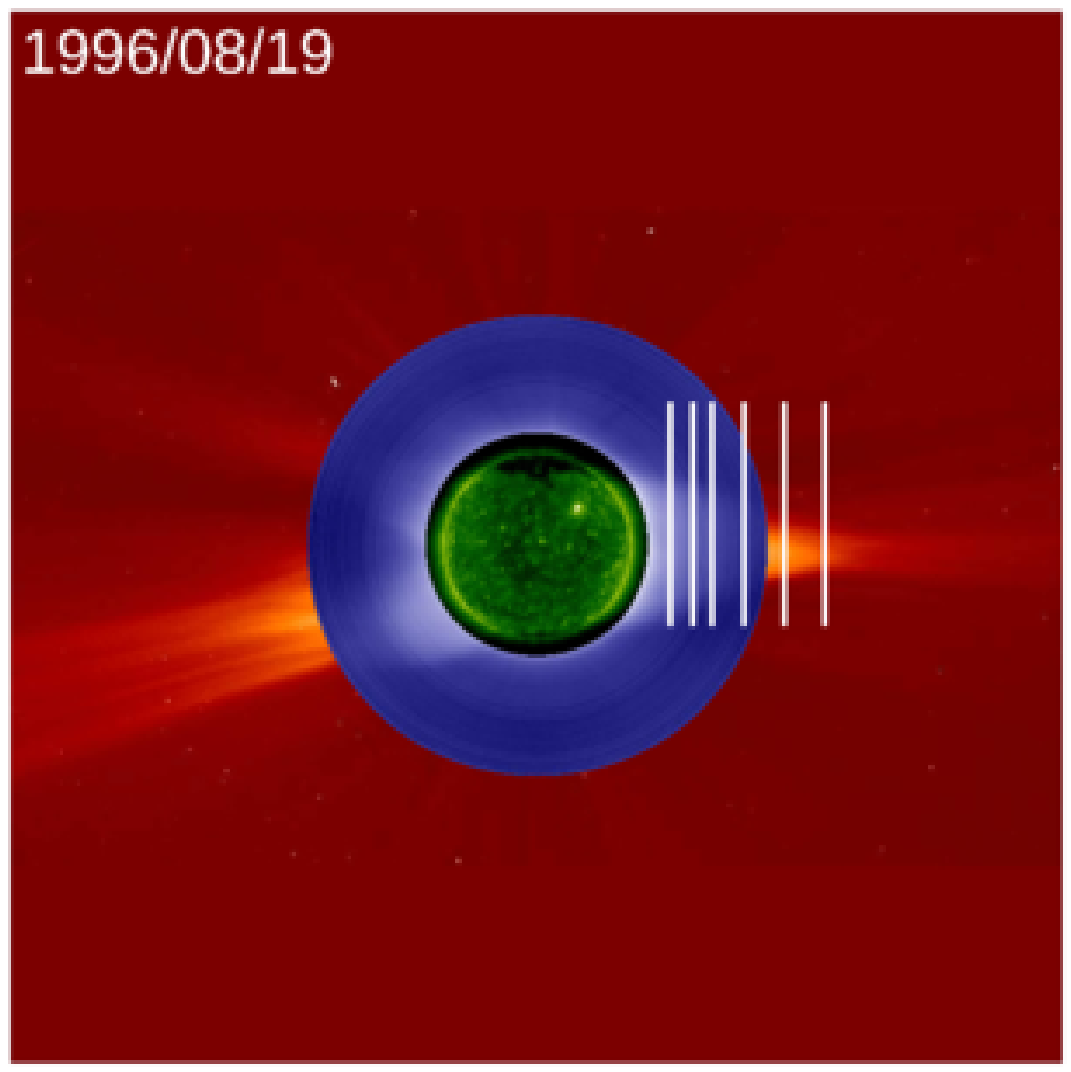}
\includegraphics[width=5.0cm,angle=0]{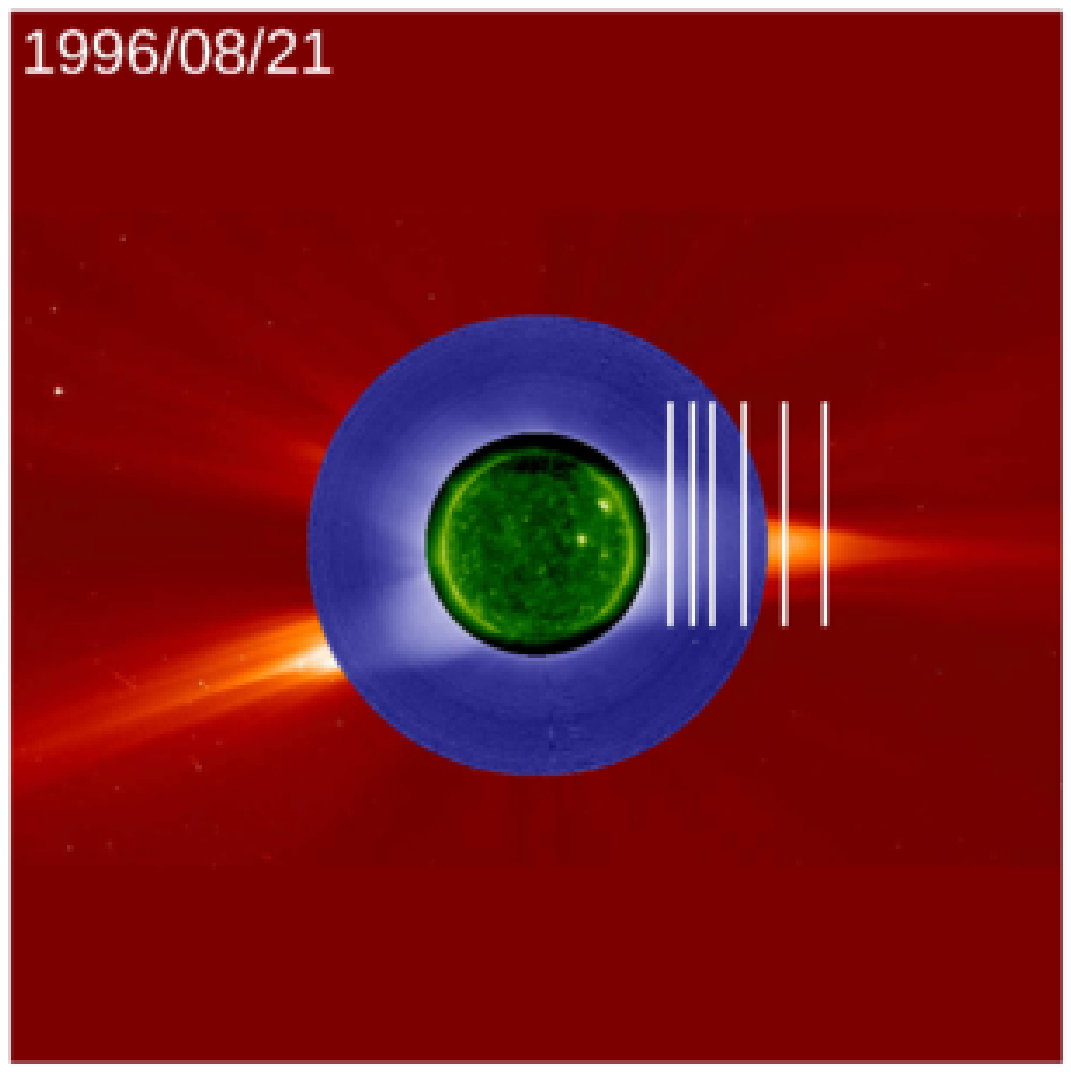}
}
\caption{Composite images including SoHO EIT 195~\AA,
HAO Mauna Loa Mark 3   and LASCO/C2 white-light
coronagraphs during times when the quiescent streamers discussed here 
were observed by SoHO UVCS, on 1996 May 27 (top left) and 1996 August 19 and 21 (bottom). 
The top right plot shows an active region streamer observed on 1996 June 1.
The location of the UVCS slit is shown.}
\label{fig:composite}
\end{figure*}
% Figure~\ref{fig:composite}

However, these UVCS  synoptic observations typically had narrow slits and 
exposure times of 500 or 1,000 seconds, generally too short to get enough signal in the 
coronal lines above 2\rsun. 
We have therefore searched the entire UVCS database of the first year of 
operation, to try and find observations of quiescent streamers 
with slit positions up to 3~\rsun\ and exposures long enough 
so we could measure the radiances of the coronal lines. 
We have visually inspected dozens of observations and 
at the end only found a few  useful datasets. 
Figure~\ref{fig:composite} shows the streamers we selected for
further analysis.
One was a quiet streamer observed on 1996 May 27. 
During the following 28-day solar rotations, on June 24, July 22, 
and August 19, the streamer became increasingly active, 
because of the emergence of the large active region 
connected to the Elephant's trunk which  we have 
mentioned. On the other hand, on the opposite side of the Sun,
the other (west) streamer was very quiescent for at least several 
days, around  August 19. Indeed this streamer has previously
been studied by several authors.
In particular by \cite{gibson99a}, where the polarized Brightness
(pB) measured by the 
HAO Mauna Loa Mark 3 and LASCO/C2 white-light coronagraphs
was used to infer  electron densities. 
We  have analyzed the observations on August 19th and 21st of the same streamer,
to see if there were significant changes. We found very little
difference in the radiances, confirming the impression 
from the LASCO/C2 images of a stable streamer. 
Figure~\ref{fig:composite} also shows one active region streamer,
observed on 1996 June 1, and discussed below.

We have used the fully-calibrated UVCS  data as available via 
the Virtual Solar Observatory. We have used the latest version 5.2
of the UVCS software (DAS) to sum the exposures and obtain the calibrated data.
We have considered the O VI primary channel,
where the O VI doublet, the \ion{H}{i} 1025~\AA\ and
the \ion{Si}{xii} 521~\AA\ (in second order) lines are present.
We have also analyzed  the redundant  O VI 
channel, where the \ion{Mg}{x} 610~\AA\ line is present in second order,
next to the hydrogen Ly $\alpha$. 

We have then averaged the line radiances over the regions
where the lines were clearly visible,  in order
to increase the signal-to-noise of the very weak coronal lines.
This clearly averages the variations that have been reported 
between the streamer cores and the boundaries, especially in terms 
of elemental abundances \citep{raymond_etal:97}.
However, we had no choice, considering that further out,  above 2~\rsun,
the streamers become very narrow and it becomes difficult to try 
and measure the core and the boundary regions. Also, 
line of sight effects mean that  boundary regions are always intermixed
with core regions.
We have carefully measured the line radiances by subtracting a 
linear background in each spectrum with custom-written software
by one of us (GDZ).  

The calibrated UVCS data assume a first-order calibration.
The UVCS radiometric calibration for first order lines
was measured before launch and tracked in flight by observing UV-bright
stars.  Such calibration was not possible for second order lines, so
we rely on laboratory measurements of the reflectivities of the mirror
materials, the efficiency of a similar grating, and sensitivity of  KBr
coated detector at the Mg X and Si XII wavelengths.  The cumulative
uncertainty on the measured radiances is difficult to estimate, but it
is probably about 30--40\%.
Now that the SOHO CDS has a reliable long-term radiometric calibration
\citep{delzanna_andretta:2015}, it is in principle possible to cross-calibrate 
UVCS against CDS, especially considering that the same spectral lines
were observed during some targeted campaigns. We have identified some useful 
datasets, but leave such a complex analysis for a future paper. 
For this paper, 
we have applied the following scaling factors to the second-order lines: 
0.14 for the \ion{Si}{xii} 521~\AA\ line  and  0.27 for the
\ion{Mg}{x} 610~\AA\ line in the redundant channel. 
These factors have been widely used in the past literature and 
an uncertainty of 40\% is reasonable for the purposes of the present paper.

\begin{figure}[htbp]
 \centerline{\includegraphics[width=7.0cm, angle=90]{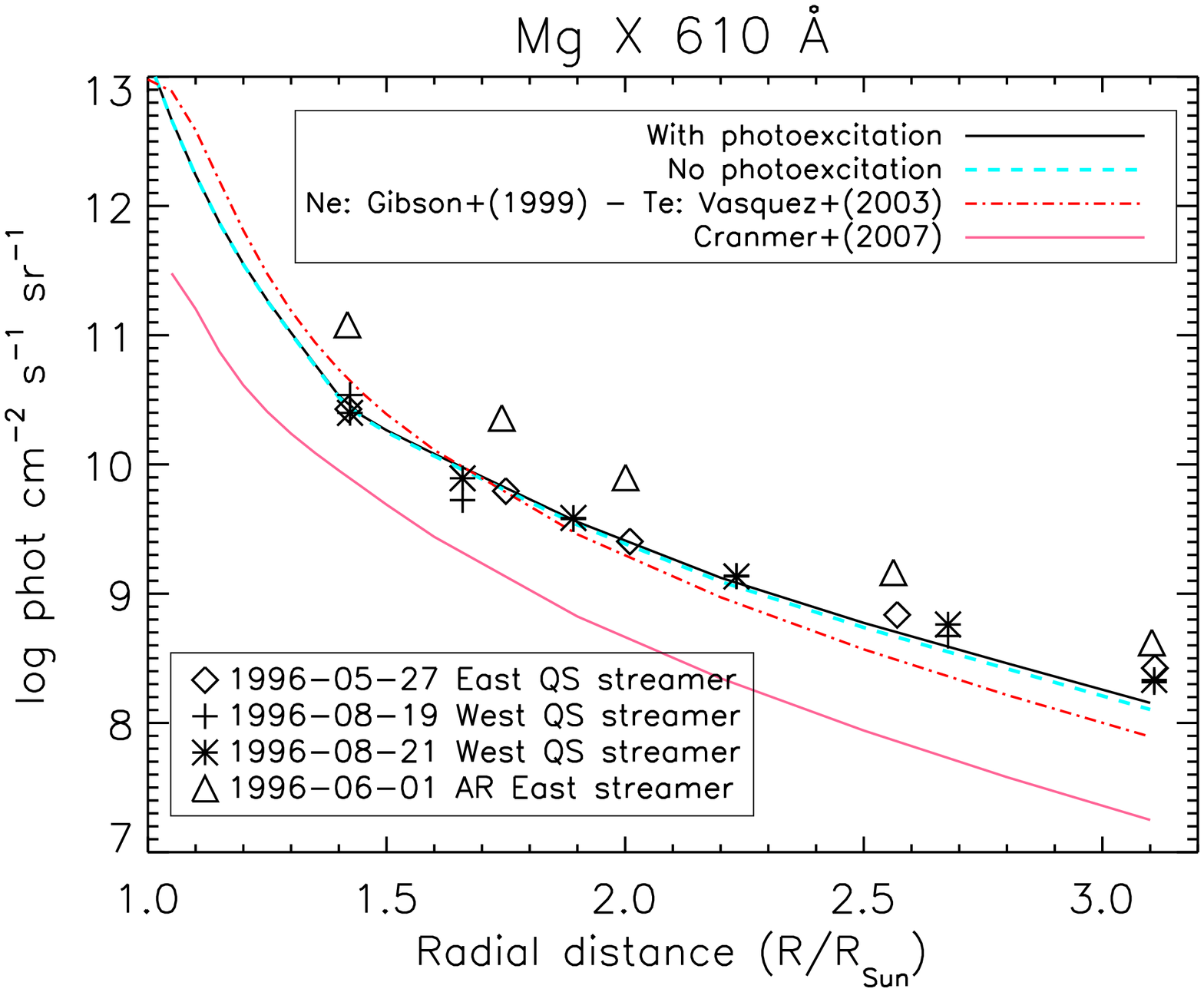}}
\caption{The radiance of the \ion{Mg}{x} 610~\AA\ line
 as a function of the radial distance, for three quiet Sun (QS) observations and one active region (AR).
The lines are the predicted radiances.}
\label{fig:mg_10}
\end{figure}

\begin{figure}[htbp]
 \centerline{\includegraphics[width=7.0cm, angle=90]{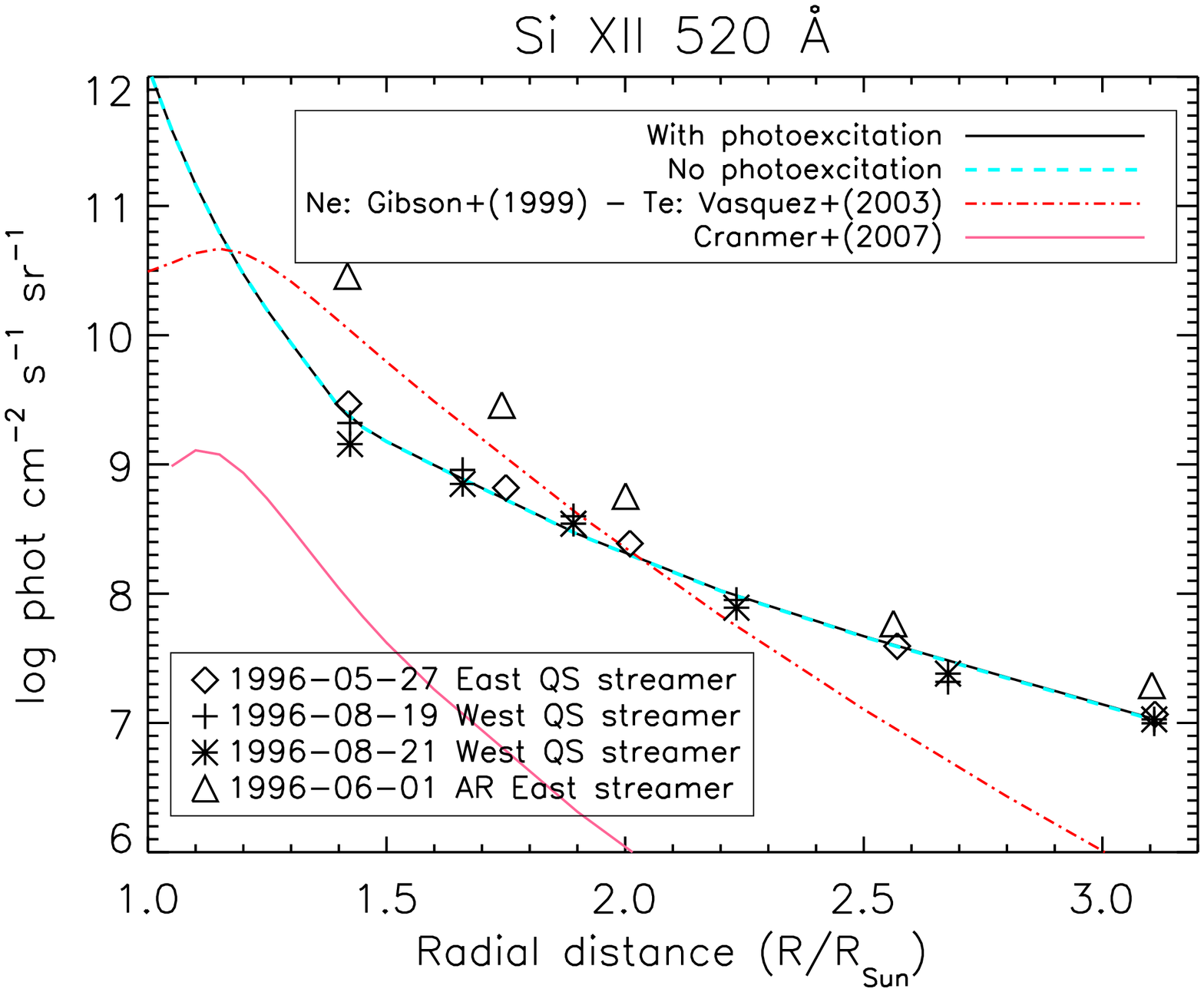}}
\caption{The radiance of the \ion{Si}{xii} 521~\AA\ line 
 as a function of the radial distance for three quiet Sun (QS) observations and one active region (AR).
The lines are the predicted radiances.}
\label{fig:si_12}
\end{figure}

The measured radiances are shown in 
Figures~\ref{fig:mg_10},\ref{fig:si_12}.
The differences between the radiances of the west streamer,
 observed on the 19th and 21st, are small. 
What is remarkable is the agreement in the radiances of the  east streamer,
indicating very little differences between the two streamers during this quiet
period. 
It is also remarkable that the radiances of the two coronal lines
decrease by only about two orders of magnitude between 1.5 and 3.1~\rsun.

% Figure~\ref{fig:mg_10}

% Figure~\ref{fig:si_12}

\section{Modeling the UVCS observations of the quiet Sun streamers and use them for COSIE-C}

For our analysis and simulations we used the atomic data from 
CHIANTI\footnote{\url{http://www.chiantidatabase.org/}}  \citep{dere_etal:97}
in its version 8 \citep{delzanna_chianti_v8}.

Given the simple structure of the streamers, we adopted a
similar approach as that one we adopted in  \cite{andretta_etal:2012},
where off-limb radiances of a selection of coronal lines and 
the He II resonance line were modeled. 
The approach is very common in the literature: 
it consists of  adopting a uniform radial  dependence of the 
electron density and temperature, estimating the line emissivities, and
then integrating  along each line of sight assuming cylindrical symmetry.
In our simulations, we calculated line emissivities up to 10~\rsun,
then performed the integrations until a radial distance of 3.1~\rsun.
For each line of sight, we integrated up to a distance of 9~\rsun,
although the main contribution to the emission comes from the 
region closest to the Sun.
This simple approach works very well in reproducing the radiances
of the collisional lines, which is the main aim of the present modeling.
This approach is also justified by several considerations,
as discussed by \cite{frazin_etal:2003}, where UVCS observations 
of another quiescent  streamer were analyzed.
However, we are exploring various other geometrical 
and density/temperature distributions for a future paper, as the 
current model is unable to reproduce the observations of the radiatively 
excited lines, see below.

Since version 4 of the CHIANTI database \citep{young_etal:03}, it is possible to 
introduce photoexcitation into the modeling of line emissivities. 
We used such an option with a few modifications by one of us (GDZ) to the software.
The model assumes a  uniform  source of disk radiation, and
calculates the excitation rate at each distance  by including a dilution factor, which 
depends on the distance from the Sun. 
We originally provided an input quiet Sun spectrum, covering all 
wavelengths from the X-rays to the infrared. However,  the electron densities
of our models are so low that the `coronal model' approximation holds: 
all the population is in the ground states and the metastable levels are not
populated. Therefore, only the disk photoexcitation at 
the wavelength of the spectral line of interest is needed.  
In the results shown here, we have assumed that the disk radiation is 
monochromatic and has the wavelength of the spectral line.

However, we have also calculated the photo-pumping by providing a disk emission 
profile and an absorption one, to study Doppler dimming effects.
Given the large Doppler widths of the coronal ions, the disk emission is 
totally absorbed in the absence of strong outflows, so the two approaches are 
fully equivalent. 

 Doppler dimming effects can become important, if strong outflows are present
\citep[see, e.g.][]{noci_etal:1987}.
In general, when outflows are present, the photo-pumping disk radiation
becomes Doppler-shifted so the  photoexcitation (and consequently the 
line intensity) decreases, although photo-pumping from nearby lines
can occur, as in the O VI case, also discussed by \cite{noci_etal:1987}.
There is ample observational evidence that quiescent streamer axes 
are characterized by negligible outflows (at most a few km/s) for the distances
we are concerned here 
\citep[see, e.g.][]{strachan_etal:2002,frazin_etal:2003,uzzo_etal:2006,noci_gavryuseva:2007,spadaro_etal:2007}.

Other assumptions we have made are that lines are optically thin,
and that the electron temperatures are the same as the proton and ion temperatures,
and the plasma is in ionization equilibrium. 
 That the electron and proton 
temperatures should be nearly equal is supported by the observation that 
the width of the  H I Lyman $\alpha$ line is close to the ionization 
temperature as measured by e.g. line ratios. The ionization temperature is equal 
to the electron one, assuming equilibrium, and the temperature of the neutral 
hydrogen must be close to the proton temperature. 
The assumption of ionization equilibrium is also common and is justified by the 
absence of any measurable flows, and the relatively high densities.
This assumption is clearly untenable in coronal hole regions, 
 where strong outflows of hundreds of km/s and much lower 
densities than in our QS streamers  have been routinely measured by UVCS. 
We note that \cite{shen_etal:2017}  computed the effects of time-dependent ionization
 in a pseudo-streamer out to 3~\rsun\  based on an MHD model.
  For that particular model, time-dependent ionization increases
 the Mg X intensity because the temperature declines with height.

Having established the method, a further problem has
been how to choose which parameters  one should adopt.
We discuss below disk radiation, electron densities, 
elemental abundances and electron  temperatures.

The accurate assessment of the disk radiation is a non-trivial issue, as different
 instruments have provided different measurements, and as radiances of all the lines
we are concerned here have a strong variation with the solar activity 
\cite[see the discussion in][]{delzanna_andretta:2015}.
For the  \ion{Mg}{x} 610~\AA\ line we have adopted a value of 110 
erg cm$^{-2}$ s$^{-1}$ sr$^{-1}$
from OSO observations, and 
for the  \ion{Si}{xii} 521~\AA\ line 19 erg cm$^{-2}$ s$^{-1}$ sr$^{-1}$. 
However, we found that  photoexcitation
for these two coronal lines has a totally negligible effect on the line intensities
for the present model.

 As discussed in \cite{andretta_etal:2012},
there is a wide range of density and temperature models for  
quiescent streamers, each of them producing very different results.
The beauty of the EUV/UV spectroscopy provided by UVCS is that 
the observed spectral lines are extremely sensitive to any
small variation in these parameters, hence provide an 
excellent test to these models.

The parameter that is better measured is the electron density.
During the first two years of the SoHO mission, one of us (GDZ) obtained
a large number of off-limb observations and monitored the off-limb
density using line ratios \citep{delzanna_thesis99,fludra99b}.
The above-mentioned near-simultaneous measurements of the pB
produced densities in relative good agreement with those measured by CDS, 
for the August 19 west streamer \citep{gibson99a}.
We therefore used as a baseline for our modelling these densities,
although some minor changes, shown in Figure~\ref{fig:ne_te}, were applied to improve 
agreement in the radiances of the  \ion{Mg}{x}  and \ion{Si}{xii} observed by UVCS.
In the same figure we also show the modeled densities for an equatorial 
quiet Sun by \cite{cranmer_etal:2007}, which have been widely  used in the literature.
Note that the paper is mostly focusing on modelling coronal holes, but
also provides other models. We have used the data file 
{\em cvb07\_equator.dat} for an  equatorial streamer at solar minimum.

We note, however, that an analysis of the same streamer by 
\cite{antonucci_etal:2005}, based on the ratios of the two O VI lines, 
produced significantly lower densities. 
That some differences in the densities occur should not be surprising.
For example, the Thompson scattered Pb signal is proportional to the 
density, while the collisionally-dominated coronal line intensities are
proportional to the square of the density. 
Therefore, if the plasma is not uniformly distributed in density, 
some changes could occur. However, the issue is more complex, as 
we briefly describe below.

Another parameter which affects the modelling is the elemental abundance.
It is well-known  that remote-sensing observations
of the solar corona, from the visible to the X-rays,
have shown that coronal abundances can be different 
from the photospheric ones, and that correlations
exist between the  abundance of an element   and its 
first ionization potential (FIP).
The low-FIP ($\leq$ 10 eV) elements are more abundant 
than the high-FIP ones, relative to their  photospheric values.
On the other hand, usually the relative abundances among 
elements with similar FIP do not vary. 
Similar FIP bias variations are observed in-situ in
 the slow solar wind. 
See, e.g. the \textit{Living Reviews} by \cite{laming:2015} 
and \cite{delzanna_mason:2018} for  details.

Since the COSIE band is dominated by lines from coronal iron
ions, and we use the  \ion{Si}{xii} and \ion{Mg}{x}  radiances
measured by UVCS for the estimates, the choice  of the 
elemental abundances is irrelevant, considering that
Si, Mg, and Fe are all low-FIP elements, so their relative
abundances are not expected to vary much. 

For our modelling, we took the simplest approach.
Considering that we averaged over the entire
sections of the streamers, we are not able to assess differences 
between the core and the outer parts of the streamers, although the 
emission primarily comes from the brightest regions, the legs. 
We therefore took as a baseline the photospheric abundances of  \cite{asplund_etal:09},
considering that there is evidence (although still debated in the 
literature) that the inner quiescent corona (around 1 MK) has nearly photospheric abundances,
within a factor of two. This has been obtained by comparing  S and Fe coronal lines
observed by Hinode EIS \citep{delzanna:12_atlas}, and a wide range of elements in a recent 
reanalysis of off-limb SOHO SUMER observations \citep{delzanna_deluca:2017}.

The parameter that is virtually unknown is the electron temperature. 
\cite{gibson99a} obtained the temperatures
shown in Figure~\ref{fig:ne_te} assuming
an hydrostatic atmosphere. As shown below, they  produce radiances in the coronal lines
that decrease with radial distance far more than  observations,
so we have discarded them. 
A much better model is the semi-empirical one developed by 
\cite{vasquez_etal:2003}. It was based on the widths of the 
H I Lyman $\alpha$ line, measured by UVCS, with the 
assumption that the electron and proton temperatures are the 
same in the streamers, where the plasma is expected to be in thermal equilibrium.
However, after some tests, also this model was unable to reproduce our UVCS observations. 
Another temperature profile shown in 
the same figure is the  modeled one for an equatorial 
quiet Sun by \cite{cranmer_etal:2007}.

We would like to point out  that we are not interested here in the real electron temperature,
but rather in the temperature which, assuming ionization
equilibrium, does fit the observed radiances in the coronal lines. 
This is in practice  a  ionization temperature. 
It is well-known  \citep[see, e.g.][]{delzanna_thesis99,fludra99b} that 
the ratio of the \ion{Si}{xii} and \ion{Mg}{x} Li-like lines
is an excellent temperature diagnostic which is not
 affected by variations in density nor in elemental abundances
(as both elements have a  low-FIP).

We have measured the ionization temperature using the UVCS measurements and found it to 
be remarkably constant around log $T$ [K] =6.15. 
 Any temperature decrease would have 
a dramatic effect on the hotter  \ion{Si}{xii} line, which would decrease
significantly.  We have therefore assumed this 
temperature, keeping in mind that the actual value is dependent on the 
accuracy of the relative calibration of the two UVCS channels where the two lines
are observed.
It interesting to note that in the inner corona, up to 1.3~\rsun, there is
spectroscopic evidence from SOHO CDS 
\citep[see, e.g.][]{delzanna_thesis99, andretta_etal:2012} and SOHO SUMER 
\citep[see, e.g.][]{landi_feldman:2003} that the ionization temperature 
is normally nearly constant. 
We provide below  further evidence from one Hinode EIS observation that the 
 ionization temperature is nearly constant up to 1.5~\rsun.
Any variation of the density and temperature above 3~\rsun\ does not 
significantly affect the modelling.

% Beyond  3~\rsun, we assumed an exponential 
%decay to reach a value of log $T$ [K] = 5.9 at 10~\rsun, following
%\cite{vasquez_etal:2003}.

The predicted intensities  are shown in the plots, together with the measured 
radiances. Excellent agreement is found for the  \ion{Si}{xii} and \ion{Mg}{x} 
coronal lines.

\subsection{On the radiatively-excited lines and the chemical abundances}

As the \ion{H}{i} Lyman $\beta$  1025~\AA\  line is largely collisional
(at least at closer distance),
we have also predicted its radiances, which allow us to measure the absolute
abundances of Mg and Si in these observations.

One of the many stunning results from UVCS was the discovery 
of very unusual abundances in  the cores of quiescent streamers, with 
high-FIP elements such as O being depleted by about an order of magnitude,
compared to its photospheric value \citep[see e.g.][ and following papers]{raymond_etal:97}. 
The low-FIP elements such as Si and Mg also appeared depleted,
 but  by not as much.
These abundance measurements have usually been obtained by comparing the 
line radiances with the radiance of the H I Lyman $\alpha$,
with some assumptions and modelling. In particular, by considering 
separately the collisional and radiative component of the lines. 
 Such low  abundances are anomalous,  as they are not observed in the 
lower corona, though they are sometimes    seen in the solar wind 
\citep{weberg_etal:2012,weberg_etal:2015}.

A possible interpretation is mass-dependent settling, which was observed
with  SOHO SUMER \citep[see ][]{feldman_etal:98b}.
Furthermore,   abundances have been found to vary
spatially not only across but also along streamers. 
 \cite{antonucci_etal:2006} analyzed UVCS observations
of  the H I and the  O VI lines and found a variation in the 
O abundance in the regions surrounding the quiescent solar minimum streamers.
\cite{uzzo_etal:2003} analyzed the UVCS observations of the 
 streamer belt from 1996 June 1 to August 5 to study the variation
of the elemental composition. 
Large variations between the streamers cores and legs were 
confirmed, as well as variations between streamers.

 For the \ion{H}{i} Lyman $\beta$  1025~\AA\  line   we used a  disk radiance of 
800 (ergs cm$^{-2}$  s$^{-1}$ sr$^{-1}$) as measured by UVCS \citep{raymond_etal:97},
noting that it is very close to the SUMER QS measurement during the same period of low
solar activity reported by \cite{wilhelm_etal:98a} of 766.
There are no significant center-to-limb changes 
in the H I Lyman $\alpha$ line, so we have assumed that the 
radiance of the Lyman $\beta$  line is constant across the disk.

\begin{figure}[htbp]
 \centerline{\includegraphics[width=7.0cm, angle=90]{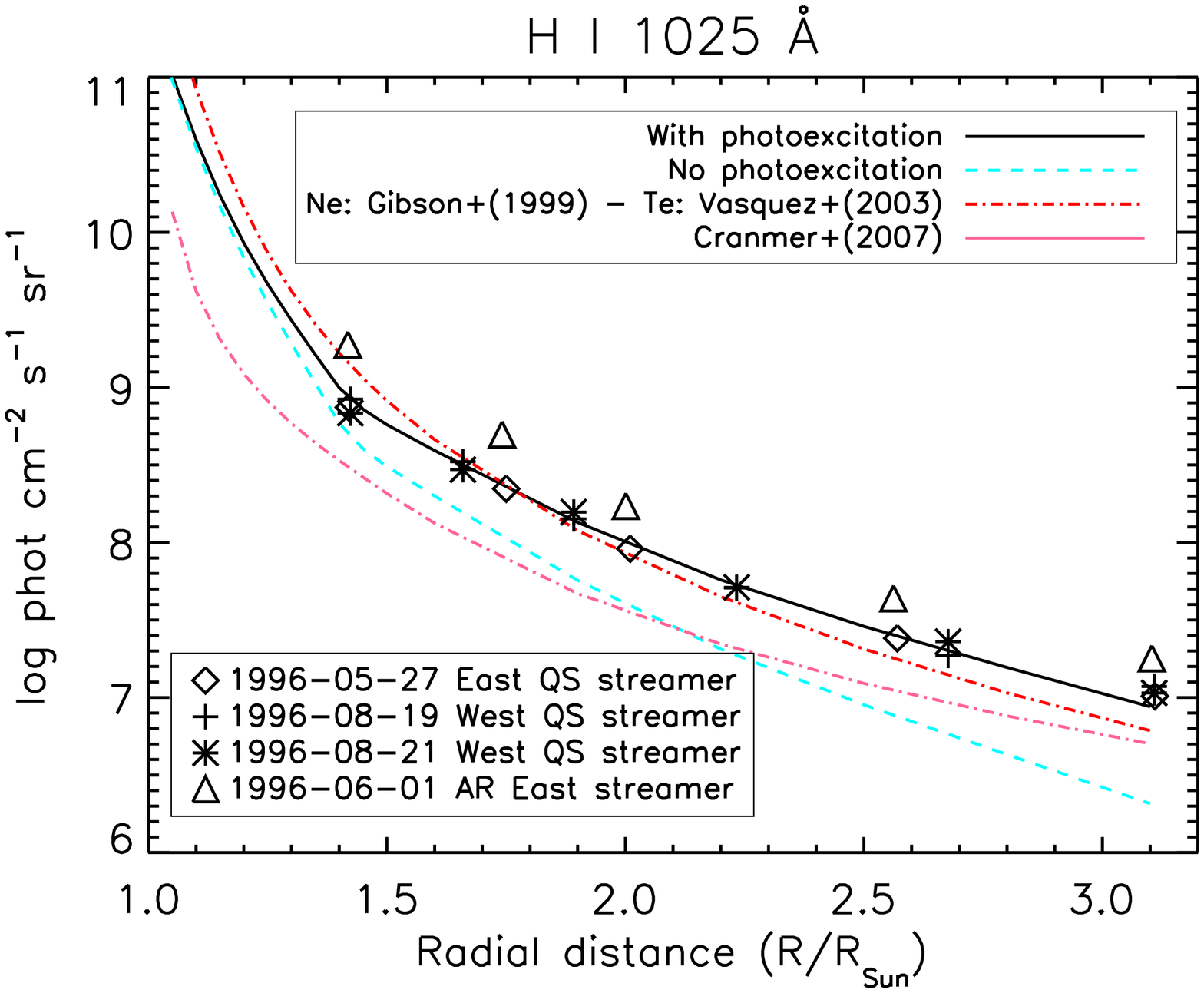}}
\caption{The radiance of the \ion{H}{i} 1025~\AA\  line
 as a function of the radial distance for three quiet Sun (QS) observations and one active region (AR).
The lines are the predicted radiances.}
\label{fig:lyb}
\end{figure}
% Figure~\ref{fig:lyb}

 Figure~\ref{fig:lyb} shows 
that its predicted radiances are very close to the 
observed ones indicates  that the Mg and Si abundances are close to photospheric.

We note that \cite{raymond_etal:97} found that in a streamer leg
at 1.5~\rsun\ the Mg and Si abundances were lower than photospheric. 
For Mg, they found an abundance of 2.5$\times 10^{-5}$, compared to the
photospheric value of 4.0$\times 10^{-5}$ we adopted here. For Si, they found a 
value of 1.25$\times 10^{-5}$, compared to a 
photospheric value of 3.2$\times 10^{-5}$.
On the other hand, \cite{raymond_etal:97} found the Fe abundance
 to be photospheric, i.e. 3.2$\times 10^{-5}$.
We note that several iron lines were observed, and the abundances 
were estimated assuming an isothermal temperature of log $T$[K]=6.2,  obtained
from the relative abundances of the iron lines: \ion{Fe}{xiii},  \ion{Fe}{xii},  \ion{Fe}{x}.
Therefore, the results concerning the iron lines should in principle
be more reliable, although significant improvements in the atomic data
for the  \ion{Fe}{xii},  \ion{Fe}{x} forbidden lines have occurred 
\citep[see ][]{delzanna_etal:12_fe_12,delzanna_etal:12_fe_10,delzanna_etal:2014_fe_9}. 
These improved data  have been distributed in 
CHIANTI version 8 \citep{delzanna_chianti_v8}.
%,  and \ion{Si}{xii}, \ion{Si}{xi}.
Other authors (cf. \citealt{parenti_etal:2000})  have found  that, on average, 
the abundances  of low-FIP elements in quiescent streamers are nearly photospheric,
within a factor of two.

There is however a caveat to these results, i.e. the fact that 
lines such as the \ion{H}{i} Lyman $\beta$  
that are partly photo-excited are very sensitive at further distances not only to the disk radiation
and outflows (via Doppler dimming), but especially to how the electron density
is distributed along the line of sight. Their dependence would be different than
that of the collisionally-dominated lines  such as \ion{Si}{xii} and \ion{Mg}{x},
and also different than the pB in the  visible continuum. 
The results at longer distances are dependent on the model because of the 
stronger photoexcitation contribution to the  \ion{H}{i} Lyman $\beta$ line, but those
around 1.4~\rsun\ are nearly independent from the photoexcitation.

With our present simple model we were able to find agreement with the 
\ion{H}{i} Lyman $\beta$ and the coronal lines, but not with the observed
radiances of the \ion{H}{i} Lyman $\alpha$ and the  \ion{O}{vi} lines. 
Discrepancies in the  \ion{O}{vi} lines have been reported previously
\citep[see, e.g.][]{frazin_etal:2003,antonucci_etal:2005,spadaro_etal:2007}
and could be due to several factors.

 As this interesting issue is well beyond the scope of the 
present paper, we defer it to a future study.
We stress here that these issues are of  importance 
for our interpretation of UVCS observations, but are completely 
irrelevant for the main aim of the present paper.
In fact, regardless of how the UVCS coronal lines are modeled,
once they are well reproduced, we can  then predict very accurately the signal in 
COSIE-C, as described below.

\subsection{Modelling the COSIE-C coronal emission}

Modelling the coronal signal expected from COSIE-C  is straightforward. 
Having established that the coronal lines have negligible
increases due to photoexcitation up to 3~\rsun,
we have therefore used the same procedure applied to model 
the UVCS coronal lines to model the COSIE-C ones.
As the main lines contributing to the band for the quiet Sun 
case are from 
\ion{Fe}{viii}, \ion{Fe}{ix}, \ion{Fe}{x}, \ion{Fe}{xi},
\ion{Fe}{xii}, and \ion{Fe}{xiii}, 
we have only calculated the emissivities of these ions.
Clearly, if e.g. a CME will enter the field of view, 
 more counts will be produced not only because of 
the higher densities, but also because of the contributions
of other cooler lines in the band.
As Fe is a low-FIP element as Mg and Si, and as these iron 
ions are formed at the same temperatures as the 
 \ion{Si}{xii} and \ion{Mg}{x}  lines, the estimate should be accurate. 

\begin{figure}[htbp]
 \centerline{\includegraphics[width=7.0cm, angle=90]{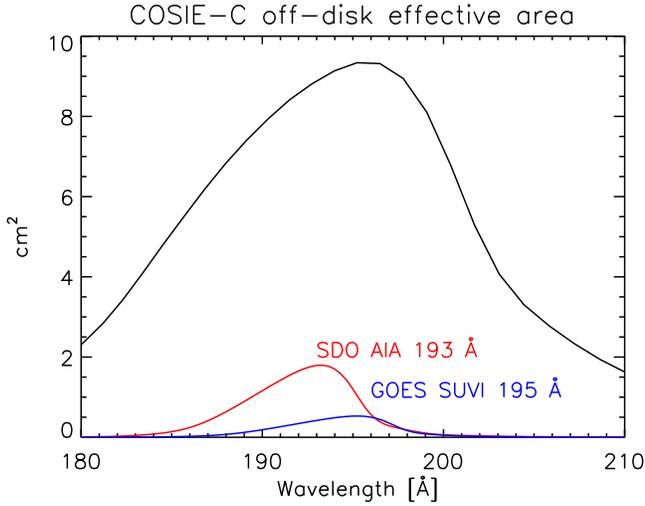}}
\caption{The estimated COSIE-C effective area, compared to those of the 
SDO AIA 193~\AA\ and GOES SUVI 195~\AA.}
\label{fig:cosie_ea}
\end{figure}
% Figure~\ref{fig:cosie_ea}

We have then folded the expected radiances $I_{\rm r}$ (integrated along the line of sight)
with the response of the 
COSIE-C instrument to estimate the signal $S$  (data numbers per second, DN/s) for a 
COSIE-C pixel  as follows:

\begin{equation}
S =  I_{\rm r} \, A_{\rm e} \, {12398.5 \over 3.65 \, \lambda \, G } \, \Omega 
\end{equation}
\noindent
where the  terms convert the number 
of electrons produced in the CCD by a photon of wavelength $\lambda$ (\AA)
into data numbers DN ($G$ is the gain of the camera),  $\Omega$ is the solid 
angle of one COSIE  pixel (3.1\arcsec $\times $ 3.1\arcsec), and  
$A_{\rm e}$ is the effective area:

\begin{equation}
 A_{\rm e} = \, A\,  T_{\rm Al} \, R_{\rm fm} \, R_{\rm m} \, T_{\rm fp} \,{\rm QE} \,
\end{equation}
\noindent
where $A$ is the geometrical area of the telescope
 (110 cm$^2$ with the current design),
$T_{\rm Al}$ is the transmission of the Al front filter,
$R_{\rm fm}$ is the reflectivity of the front folding mirror, 
$R_{\rm m}$ is the reflectivity of the main mirror, 
 $T_{\rm fp}$ is the  transmission of the internal focal plane filter, 
and QE is the quantum efficiency of the detector.

As a guideline, we have used for $T_{\rm Al}$ a standard 
Al front filter with an anti-oxidant layer, similar to the 
Hinode EIS flight front filter, but with  an improved 
supporting mesh (95\% transmission), instead of the 
one used for EIS (85\% transmission).
 The improved mesh has  been flown several times, e.g. for the Hi-C rocket
flights \citep[see, e.g.][]{kobayashi_etal:2014}. 
The total transmission is nearly 60\%.
For $R_{\rm fm}$ we have used an estimate for a Zr/Al mirror, resulting 
in about a 65\% reflectivity across the wavelength range.
For $R_{\rm m}$ we have used the measured reflectivity of the Si/Mo multilayer
used for the flight Hinode EIS short-wavelength channel, which has a peak
of 35\% around 195\%. 
For the focal plane filter we have assumed again an Al filter with a supporting mesh.  
As the guideline detector should be an improved version of the  CCD
used for Hinode EIS, we have used a  QE=0.8, 
 based on measurements of CCDs with enhanced backside treatments
(note that EIS has a lower QE of about 0.55). 
As an example, the measured QE of the AIA CCDs is between 0.8 and 0.85
at the COSIE wavelengths \citep[see][]{boerner_etal:12}.
For the gain, we have used the same
estimated for EIS:  $G$=6.3 electrons/DN. 
Note that, due to the large geometrical area, the peak values of the 
effective area  are much higher (about 9 cm$^2$) than  
previous instruments, including the SDO AIA 193~\AA\ and GOES SUVI 195~\AA,
 see Figure~\ref{fig:cosie_ea}. 
For  AIA we used the measured values, as available via SolarSoft, while for 
SUVI we used approximated values based on current understanding of the 
instrument (D. Seaton, private communication).

\begin{figure}[htbp]
 \centerline{\includegraphics[width=7.0cm, angle=90]{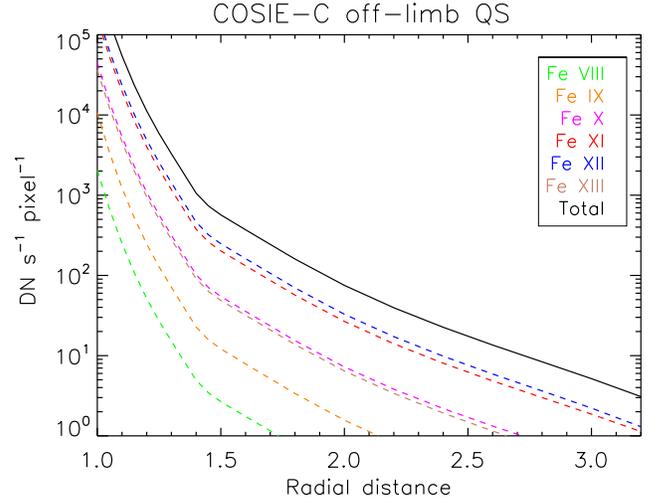}}
\caption{The estimated total count rates (per pixel) in the 
COSIE-C broad-band, 
 as a function of the radial distance, for the quiet Sun,
using the UVCS model. 
The total rates are shown, as well as those produced by the main ions. }
\label{fig:cosie_qs}
\end{figure}
% Figure~\ref{fig:cosie_qs}

\begin{figure}[htbp]
 \centerline{\includegraphics[width=7.0cm, angle=90]{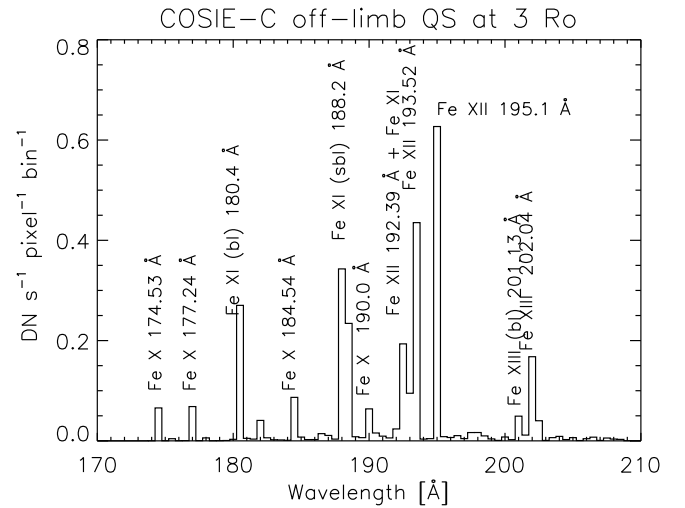}}
\caption{COSIE-C predicted spectrum at 3\rsun.}
\label{fig:cosie_qs_spectrum}
\end{figure}
% Figure~\ref{fig:cosie_qs_spectrum}

The resulting total count rates expected for off-limb 
quiet Sun observations  are shown in Figure~\ref{fig:cosie_qs}. 
As expected, the signal is so high close to the limb that 
a filter will have to be used, together with very short exposure times,
which will allow unprecedented high-cadence observations, useful to 
study fast dynamics and waves. 
Very high cadence of tens of seconds could in principle be achieved
 even at 3~\rsun\ in bright streamers and transient solar wind structures,
 depending on the stray light levels (see discussion in Section 5).

Figure~\ref{fig:cosie_qs}  also shows a breakdown of the contributions from 
the main ions. As expected, the dominating contributions are from 
 \ion{Fe}{xi}, and \ion{Fe}{xii}, i.e. ions that are formed
at similar temperatures. 
 In other words, although COSIE-C is a 
broad-band instrument, it is not as multithermal as most of the 
SDO AIA bands, although extra contributions 
from much cooler (\ion{O}{v}) or hotter (\ion{Ca}{xvii} and \ion{Fe}{xxiv}) lines 
could occasionally be present.
 As shown e.g. in \cite{delzanna_etal:2011_aia} and 
\cite{delzanna:2013_multithermal}, 
having a very multithermal band often complicates the analysis of solar observations.

\section{QS Hinode EIS off-limb}

\begin{figure}[htbp]
% \centerline{\includegraphics[width=5.5cm, angle=90]{fe_12_overlay1.ps}}
 \centerline{\includegraphics[width=7.7cm, angle=0]{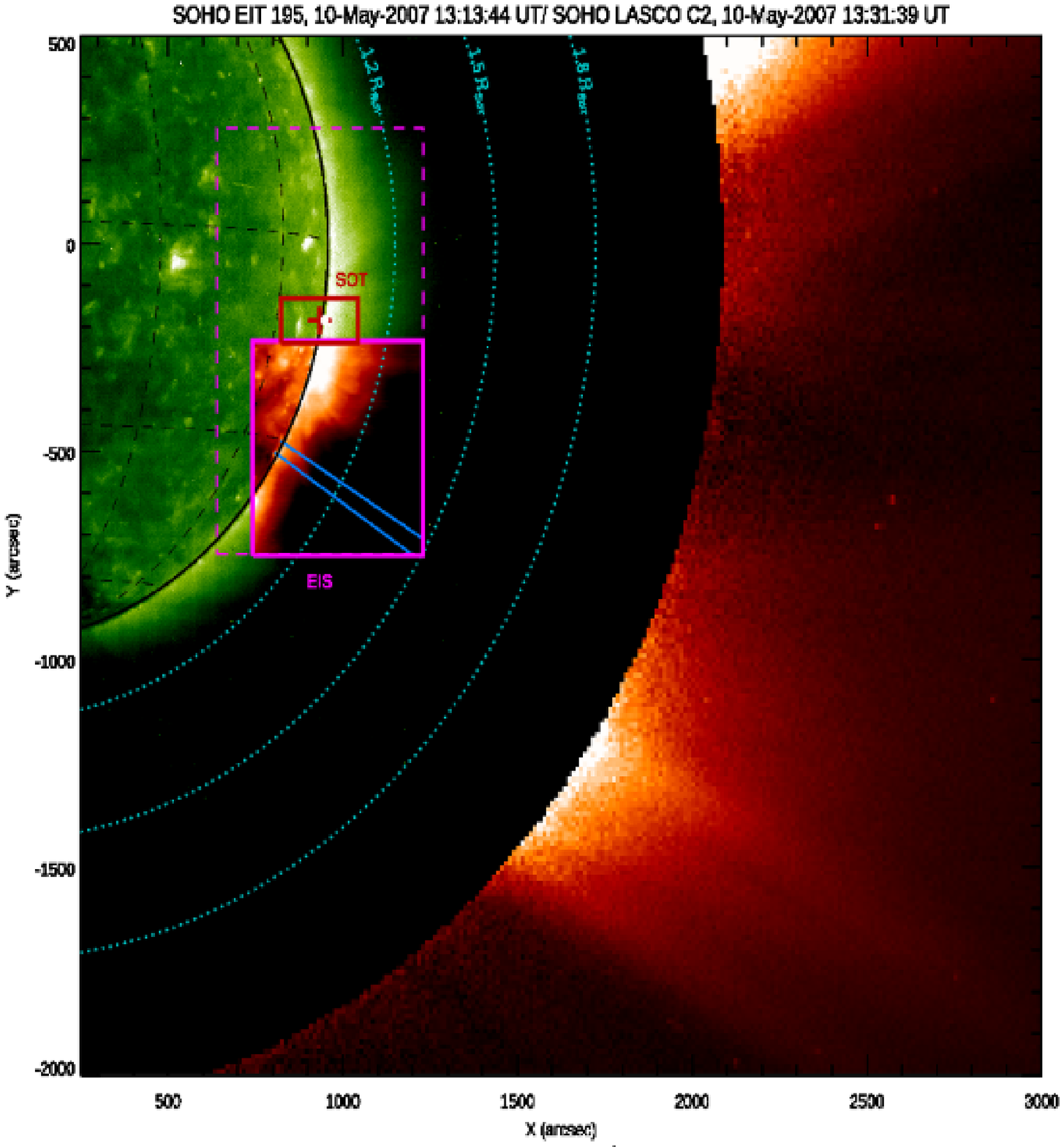}}
 \centerline{\includegraphics[width=7.cm, angle=0]{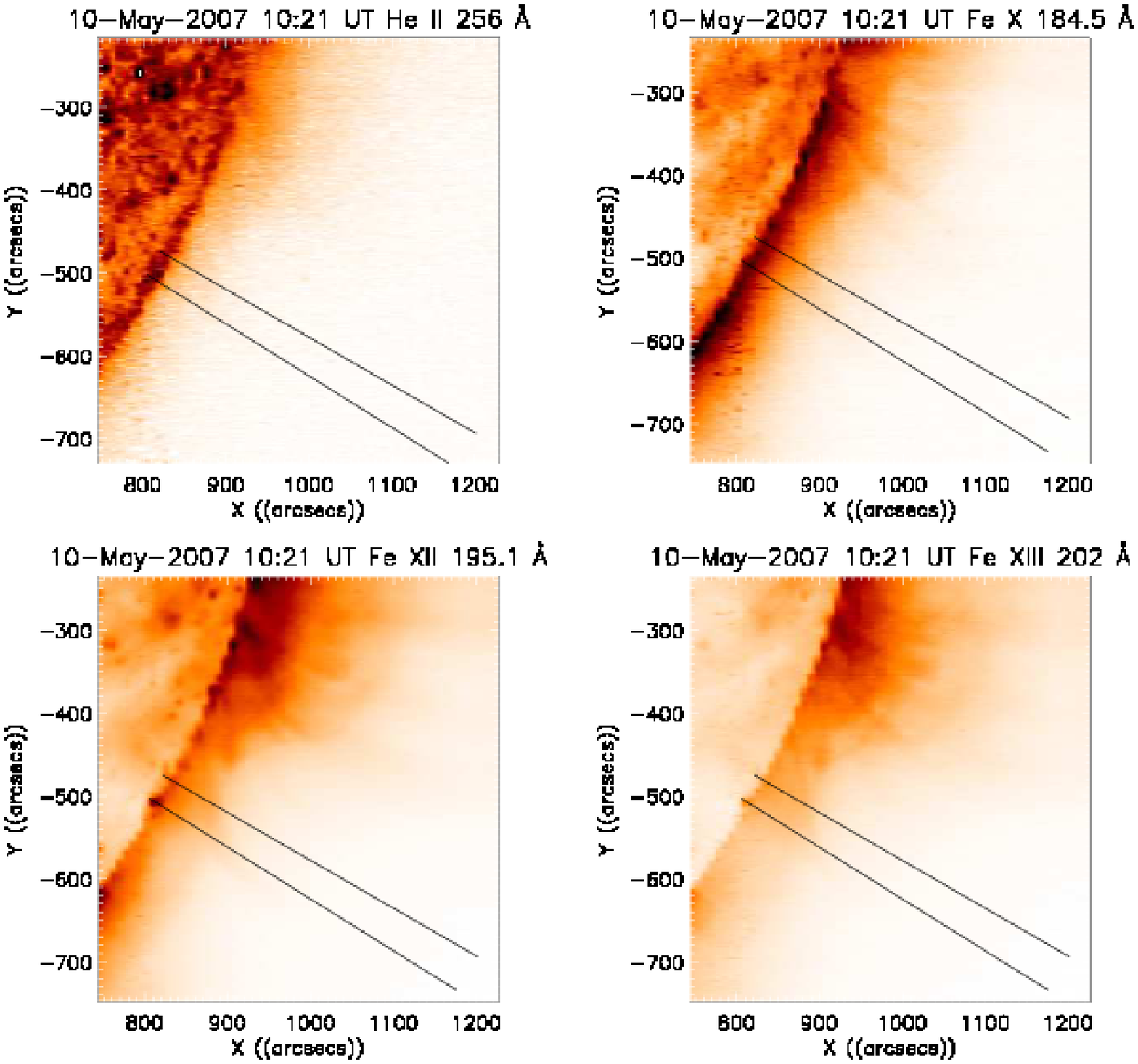}}
 \centerline{\includegraphics[width=6.5cm, angle=90]{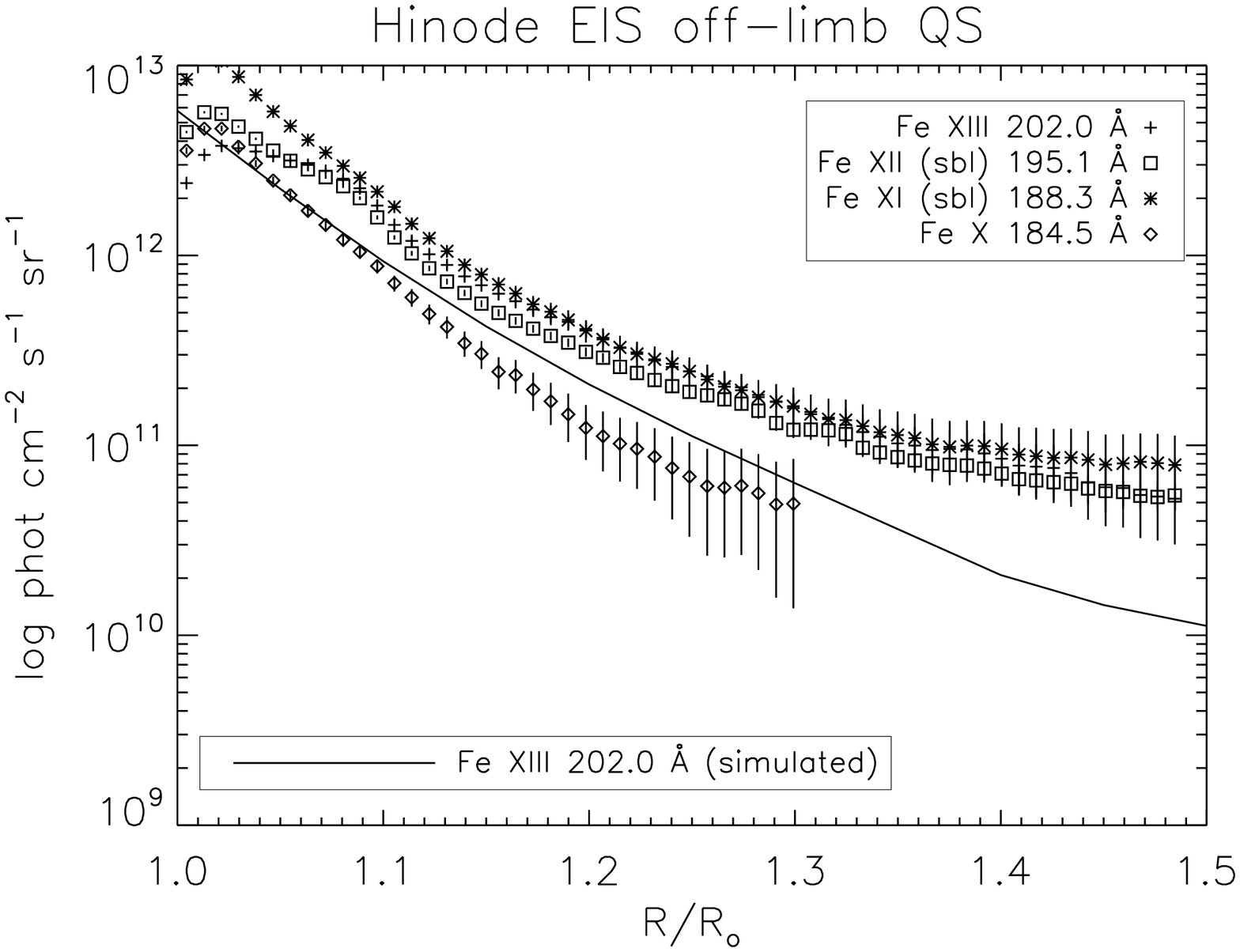}} 
\caption{From top to bottom: 1) a SOHO EIT 195~\AA\ image on 2007 May 10, 
with the Hinode EIS field of view (full possible, pink dashed rectangle;
 observed, full pink lines) and the EIS \ion{Fe}{xii} 195.1~\AA\ monochromatic
image. The blue lines indicate the radial sector used for the averaging (see text).
The SOHO LASCO C2 image is superimposed, showing that the 
selected radial sector is in  a streamer.
2) Monochromatic images (negative) in a few spectral lines, with the radial sector
indicated.
3) the observed radial variation of the radiances of the main 
coronal lines. The full line is the 
estimated radiance in the   \ion{Fe}{xiii} 202.0~\AA\ line, based on the 
modelling of the 1996 solar minimum quiet Sun streamer. }
\label{fig:eis_ol}
\end{figure}
% Figure~\ref{fig:eis_ol}

As Hinode pointing outside the solar limb is normally 
not allowed, almost all the off-limb quiet Sun EIS observations are restricted to
about 1.2~\rsun. 
There are only a few observations to greater distances, but they were in 
polar coronal holes, where the edge of the EIS slit reached about 1.4~\rsun. 
One of us (GDZ)  designed an engineering
EIS 'study' to extract spectra from the bottom half of the long EIS slit,
to reach greater distances. 
A week-long campaign  (Hinode HOP 7) was coordinated to obtain simultaneous 
 SOHO/Hinode/TRACE/STEREO observations during the SOHO-Ulysses quadrature in May 2007. 
EIS observations were obtained during May 7--10, as outlined in \cite{delzanna_etal:2009ASPC}. 
The Sun was very quiet, with the exception of an active region, which was
at the west limb around May 7, as shown in Fig.~\ref{fig:eis_ol}(top).
The Figure also shows the  field of view covered by  EIS.
The lower part of the EIS field of view  reached the significant  distance of 1.5~\rsun.
On May 10, the main part of the active region was behind the limb. 
We have selected the observation which started on that day at 10:21 UT.
The EIS  2\arcsec\ slit was moved with 8\arcsec\ jumps, to cover about 500\arcsec\
in the E-W direction. A long exposure of 60s was chosen. 
As far as we are aware, this is the only EIS observation of the quiet Sun up to 
such large distances. 

The EIS data have been processed with custom-written 
software written by GDZ and radiometrically calibrated using  the \cite{delzanna:13_eis_calib} 
analysis (see below).
Monochromatic images in a few lines are shown in  Fig.~\ref{fig:eis_ol}(middle).
The \ion{He}{ii} 256~\AA\ line was deblended from the contribution 
of a strong coronal \ion{Si}{x} using the 261~\AA\ line. 
Some emission associated with the active region is still visible in the 
north portion of the EIS field of view. Aside from a region up to  1.1~\rsun,
where some structures are present, the outer corona is featureless, although 
it bears some differences with the quiet Sun observations during the 1996 solar 
minimum presented above.

An off-limb  region about 5$^o$ wide, along the radial direction indicated in 
 Fig.~\ref{fig:eis_ol}  was chosen, to obtain 
averaged intensities as a function of the radial distance, which are shown in 
 Fig.~\ref{fig:eis_ol} (bottom).
They were obtained by first averaging  about 15-20 EIS pixels, then obtain 
the averaged spectra and then the integrated intensities. 
The uncertainties include the Poisson noise associated with the number of 
detected photons in the line and nearby pseudo-continuum, plus the read-out noise.

\begin{figure}[htbp]
 \centerline{\includegraphics[width=7.0cm, angle=90]{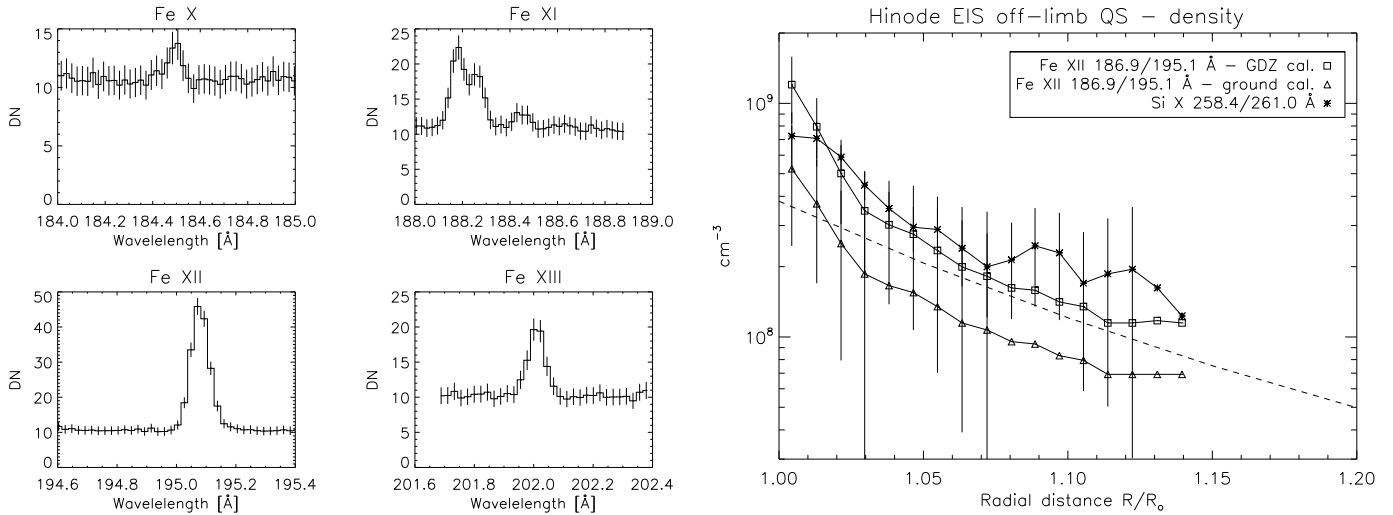}}
\caption{Hinode EIS spectra of the stronger lines at 1.5~\rsun.}
\label{fig:eis_spectra}
\end{figure}

Only a few of the strongest EIS lines had good signal at 1.5~\rsun:
 \ion{Fe}{xi} 188.2~\AA\ (a self-blend), the \ion{Fe}{xii} 192.4~\AA\ and 
195.1~\AA\ (a self-blend), and the \ion{Fe}{xiii} 202.0~\AA. The
 \ion{Fe}{x} 184.5~\AA\ line was barely visible.
Figure~\ref{fig:eis_spectra} shows averaged spectra at the larger distance,
1.5~\rsun. Note that the apparent continuum around 10 DN is actually the 
detector bias, which was not subtracted. 
All the other lines  had strong signal only up to about 1.2--1.3~\rsun.

% Figure~\ref{fig:eis_spectra}

The behavior of the cooler lines, discussed below when we consider 
stray light, indicates negligible stray light for this observation,
hence confirming that the signal in the coronal lines is real, and not
scattered light from the disk. 
 Fig.~\ref{fig:eis_ol} (bottom) shows with a full line the radiances
of the  \ion{Fe}{xiii} 202.0~\AA, as estimated using the 
1996 UVCS streamer observation. This is significantly lower than the observed 
 \ion{Fe}{xiii} 202.0~\AA\ radiance.

\begin{figure}[htbp]
  \centerline{\includegraphics[width=7.0cm, angle=90]{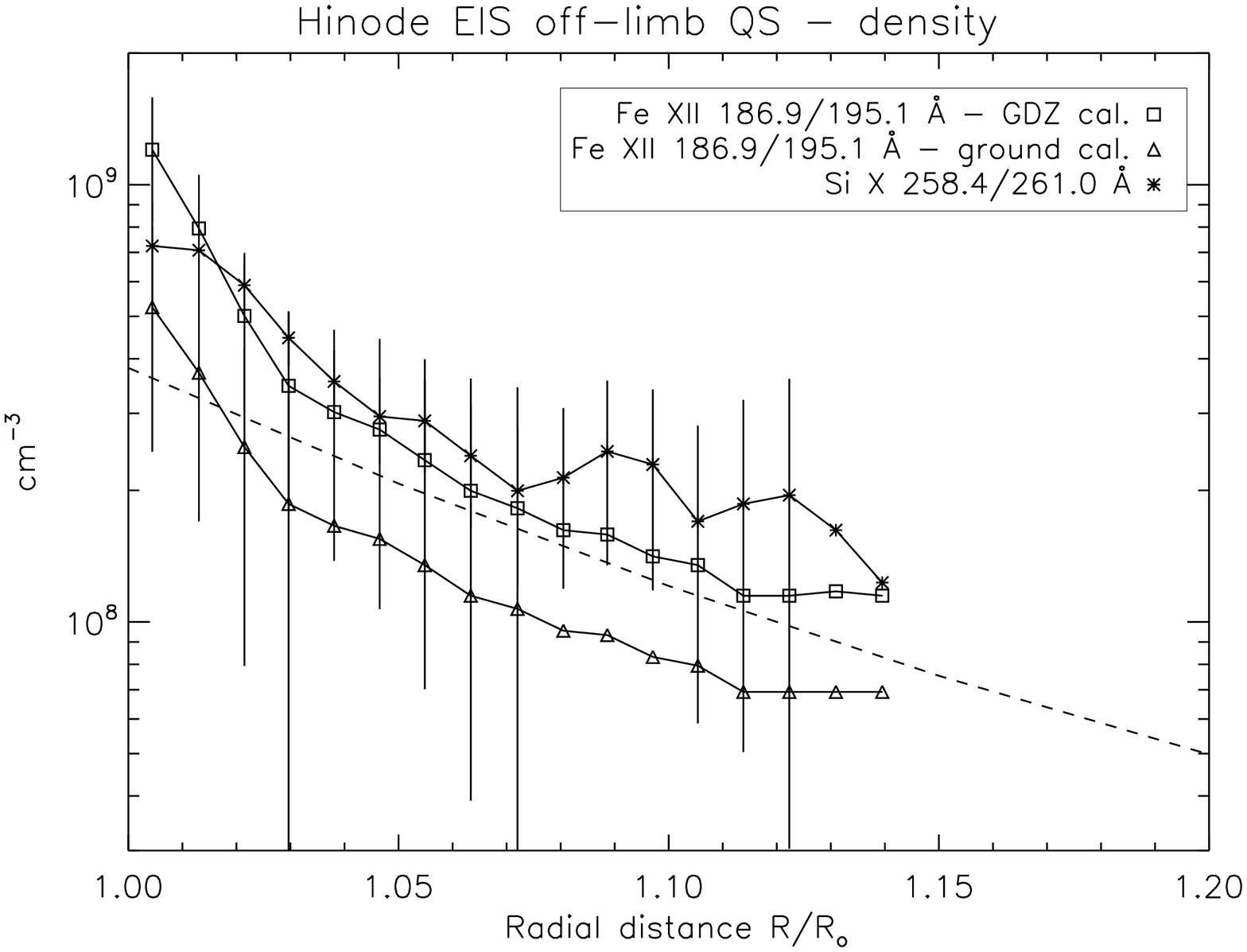}}
  \centerline{\includegraphics[width=7.0cm, angle=90]{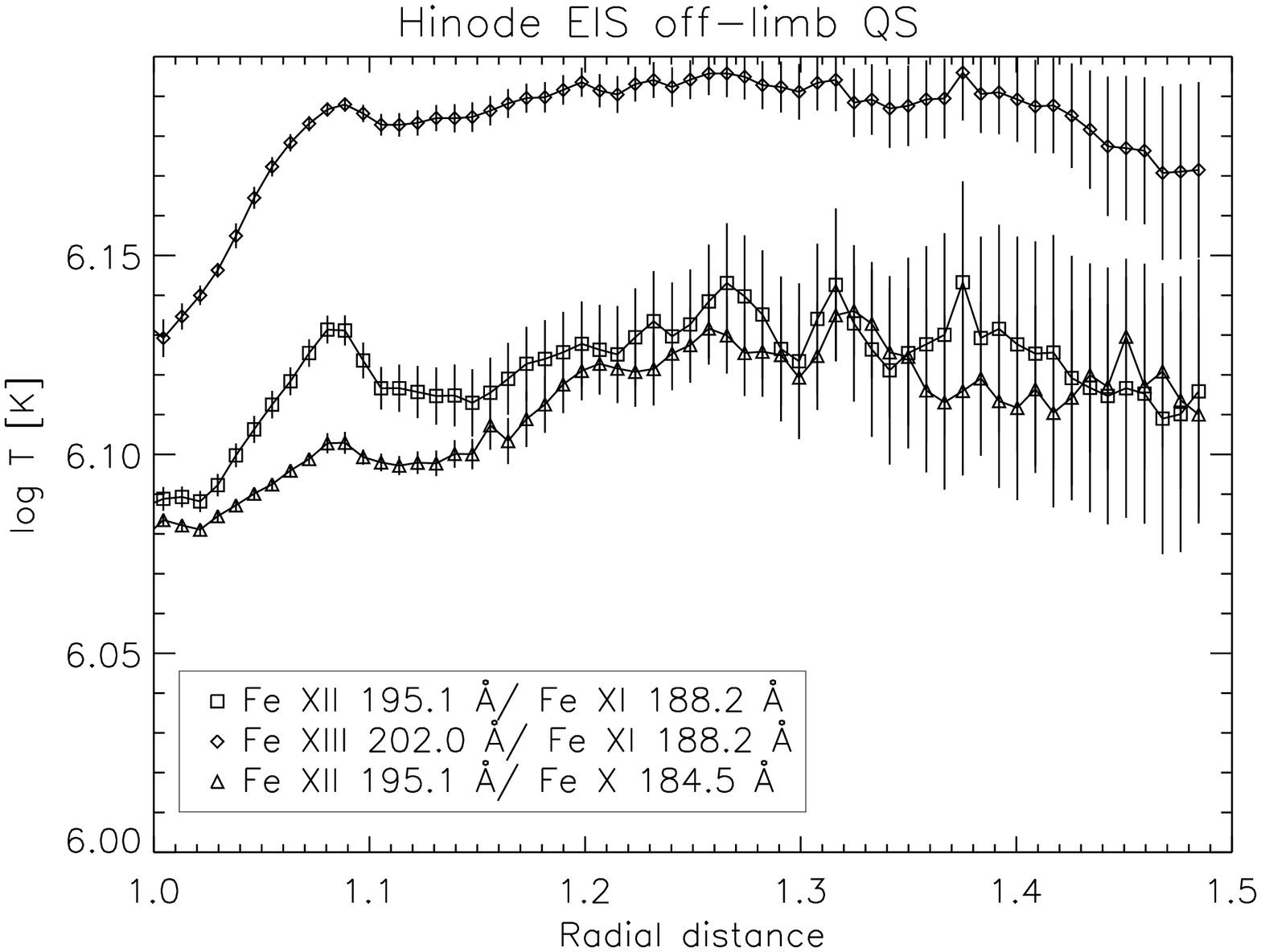}}
\caption{Averaged densities and temperatures obtained from EIS.}
\label{fig:eis_nte}
\end{figure}
% Figure~\ref{fig:eis_nte}

We have explored the possible reasons for this difference, by measuring the 
densities and temperatures from the EIS lines, which are shown in 
Figure~\ref{fig:eis_nte}.
Measuring densities is very difficult because of the low signal 
in the density-sensitive lines. 
We obtain relatively good agreement between two of the main 
density ratios available to EIS at 1--2 MK, from  \ion{Fe}{xii}
and  \ion{Si}{x} (for  a discussion of available ratios see \cite{delzanna:12_atlas})
 using the \cite{delzanna:13_eis_calib} radiometric calibration,
and not the ground calibration, where the   \ion{Fe}{xii} densities are
unreasonably low. 
The averaged electron densities of this region/streamer are slightly higher than 
those we assumed in our model, which explains some of the differences. 

Some further differences are probably related to  the temperature of the plasma.
 Figure~\ref{fig:eis_nte} (bottom) shows that we obtain an excellent
agreement in the ionization temperatures from ratios involving
\ion{Fe}{x}, \ion{Fe}{xi}, and \ion{Fe}{xii}. It is remarkable that 
these also indicate a  temperature (averaged along the line of sight)
nearly constant with radial distance, although slightly lower, 
log $T$ [K]=6.12 instead of log $T$ [K]=6.15. 
Note that this result is surely independent from any stray light.
The absolute value depends however on the accuracy of the 
ionization and recombination rates, which are constantly under revision.
On the other hand, the ratio involving the hotter  \ion{Fe}{xiii}
indicates the presence of a slightly hotter component. 
This was to be expected, considering the presence of the active region not far away.

\section{Stray light}

Stray light, i.e. the light scattered within an instrument, is an important 
quantity to be estimated for any instrument. Such estimates are notoriously 
difficult to obtain, and have not received as much attention in the literature as they should
have.  Stray light varies significantly with each instrument
(as it depends e.g.  on the micro-roughness of the optics, the filters,
 the optical paths),  
but also depends on the illumination, i.e. on where one is observing.

Out-of-field stray light is often estimated by off-pointing the spacecraft,
while in-field stray light is often estimated looking at the on-disk signal 
when part of the Sun is occulted by a planet, such as Venus or the Moon. 
The signals are often close to the noise levels, and it is always hard to assess
for the real presence of a coronal signal. 

Several EUV imagers have shown non-negligible stray light, in the wings of the PSF,
for example SoHO EIT \citep{auchere_etal:2001}, 
STEREO EUVI \citep[see, e.g.][]{shearer_etal:2012} and SWAP/PROBA2 
\citep[see, e.g.][]{halain_etal:2013,goryaev_etal:2014}. Earlier instruments had 
a lot more stray light, while the performance of more recent ones 
has improved significantly.
One earlier instrument with an excellent
performance was  the 1998  SwRI/LASP Multiple XUV Imager (MXUVI),  which 
flew on a sounding rocket for the 
SoHO EIT calibration \citep{auchere_etal:2001}. The instrument had a highly-polished (0.5~\AA\ rms) 
single mirror with a multilayer in the  171~\AA\ band, a front
filter, a light trap and baffles to reduce  scattered light. 
The detected signal  was about 1\% the disk intensity at 1.5~\rsun, and 
about  $3 \times 10^{-3}$ at 2~\rsun\ \citep[cf. Fig.~8 in ][]{auchere_etal:2001}.
Considering that the strong coronal \ion{Fe}{ix} 171~\AA\ line would 
always emit some signal in this band  even at such distances, one could argue that 
the level of stray light was for sure less than the detected signal.

Another instrument with excellent performance was the Normal Incidence 
X-ray Telescope (NIXT), which was developed at SAO and used multilayers.
This instrument paved the way for all the subsequent EUV multilayers flown in
all the solar missions.
 During a NIXT rocket flight on 1991 July 11, a knife-edge test 
using the edge of the Moon was carried out. The results indicated that the level
of scattered light was less than  $5 \times 10^{-4}$ the disk intensity 
\citep{spiller_etal:1994}.

The COSIE optical setup is unique, and a thorough assessment of the stray light will
only be possible once all the components are fabricated and measured. However, as 
all the components have significant heritage from other missions, we can provide
some comments on what would be expected, mostly based on the performances of 
SDO/AIA and Hinode/EIS.

Two of the main sources of stray light we expect are the  scattering of the mesh supporting
the two aluminum filters, and the scattering of the optics. 
The main mirror and multilayer will have performances improved over the
SDO AIA ones, and similar to those of the Hi-C instrument. The mesh will be a significant 
improvement. 
The AIA telescopes have two filters (one internal and one external)
and two mirrors, polished to a micro roughness of about 
1--2.5~\AA\ rms \citep{soufli_etal:2007}, and both coated with multilayers \citep{lemen_etal:12}.
As described in the on-line report on the AIA PSF \citep{grigis_etal:2012}, 
 by far the single biggest factor contributing to scattered light in AIA 
is the diffraction caused by the support mesh of the filters. That can be
quantified and deconvolved.  There
remains a residual from imperfect removal of that diffracted component,
and the main way to reduce  it is to use a much wider mesh. This has been 
implemented for the  Hi-C rocket flights \citep{kobayashi_etal:2014}
and will be implemented for COSIE.
The diffracted component due to the  mesh goes down from 20\% in all previous instruments,
including EIS and AIA, to about 2\%, hence will be much reduced for COSIE. 
Observations of the lunar limb during partial eclipses were used to get the scattering in
each of the AIA channels. The scattering in the EUV was estimated to be
about 10$^{^-5}$ per surface, i.e. negligible. 
The only other source of scattering we are aware of is some very wide-angle 
 scattering produced by multilayers, which  has been measured in one instance
\citep[see, e.g.][]{2010SPIE.7732E..37M}. 

\cite{gonzalez_etal:2016} estimated the AIA PSF using both lunar and Venus
occultations, however the validity of using Venus has been questioned 
\citep{afshari_etal:2016}. 
\cite{poduval_etal:2013} also used a lunar occultation to provide an estimate
of the AIA PSF. Typically, both the full-Sun deconvolution method 
 \citep{grigis_etal:2012}, and that one suggested by 
\cite{poduval_etal:2013} produce similar results, with changes in the 
on-disk intensities by 10--20\%, but mostly negligible off-limb, with a 
large amount of noise.

One could in principle estimate an upper limit on stray light directly 
from off-limb observations. However, the coronal EUV bands in AIA cannot be used
as the corona would produce some signal. The He II 304~\AA\ line has a significant 
coronal emission \citep{andretta_etal:2012} and most likely a resonantly
scattered component \citep{delaboudiniere:1999}, so it cannot be used. 
The only band perhaps usable is the 131~\AA,
as this band is dominated by cool emission, mostly from \ion{Fe}{viii} 
with some contributions from \ion{Ne}{vi} and \ion{O}{vi}  
\citep{delzanna_etal:2011_aia}. 
The signal for this band quickly reaches near off-limb
(i.e. even in a region where some solar signal should be present)
 a noise level (i.e. 1--2  DN per pixel) 
at about 8\% the average on-disk signal, with the standard 3s exposures for this band.
There are also a few off-point AIA observations which in principle 
could be used. We analyzed those of  2013 Nov 13 and 28, 
but did not find them suitable to assess stray light, as standard exposures were used. 

\begin{figure}[htbp]
  \centerline{\includegraphics[width=7.0cm, angle=90]{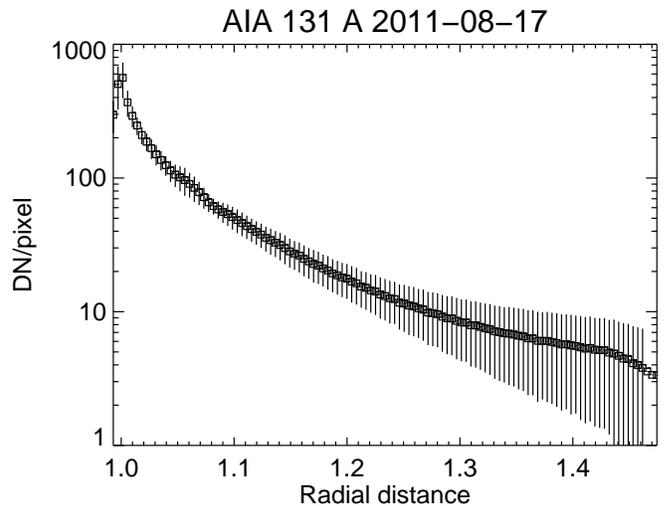}}
\caption{The observed radial fall-off in the AIA 131~\AA\ band.}
\label{fig:aia_ol}
\end{figure}

There are calibration data with longer exposures. Among the longest exposures
(32s), we analyzed those  taken on 2011 Aug 17. 
Even with these long exposures the average signal on-disk is only about 200 DN
(see Figure~\ref{fig:aia_ol}),
so if there was say a 2\% stray light at 1.5~\rsun, one should see a flattening of the 
off-limb signal towards about 4 DN. This is not  observed, as shown in 
Figure~\ref{fig:aia_ol}, although it also cannot be ruled out, given the 
low signal and noise. To obtain the points in the figure,
 we have averaged four consecutive exposures 
(using level 1 data) over a radial sector in a quiet Sun SW region. 
Solar structures are clearly
visible out to 1.2--1.3~\rsun, so some solar contribution out to 1.5~\rsun\
is still possible. 
The optical layout of the AIA telescopes is such that the FOV is annular, 
i.e. in the images there are four corners, above 1.54~\rsun, that are 
not illuminated. They can be used to assess the level or read-out noise
in the level 1 data (which have been corrected for the dark frames and 
flat-field). We found an average DN=0.7 with a standard deviation of 2.5 DN/pixel.
The 'error bars' in Figure~\ref{fig:aia_ol} are the standard deviation 
of the ensemble of points averaged at each radial distance, plus the 
 2.5 DN/pixel noise level, to provide an estimate on the level of noise
in the data.

% Figure~\ref{fig:aia_ol}

The stray light in Hinode EIS has not been studied in depth. 
Lunar occultations were used by \cite{ugarte:2010} to suggest a 2\%  level 
of stray light for the instrument. This was obtained from the residual signal
in the 195~\AA\ line, once  dark frames were removed. A broad 
distribution of DN values, centered at  5.2 and 5.8$\pm$4.1 was found,
depending on which dark frame was used.  Note that
the dark frames are highly variable in time and spatially, and have large 
DN values, of about 500.  The same procedure was applied to the strongest 
 line in the long-wavelength channel at 256~\AA, to find 
distributions  centered at -1.2$\pm$3.6  and 0.8$\pm$3.7 DN. It is unclear 
why the two results are so different, and if a 2\% stray light is real.
More importantly, it is not clear how  such on-disk observations could be used 
to estimate the actual stray light in far off-limb observations.

\cite{hahn_etal:2012} used a model to estimate the near off-limb (up to 1.3~\rsun) 
signal in coronal holes one should expect  in the 256~\AA\ line and
 found excellent agreement with the observed fall-off (cf. their Fig.~2) which was well
below the 2\% level, suggesting  negligible stray light off-limb.

\begin{figure}[htbp]
  \centerline{\includegraphics[width=7.0cm, angle=90]{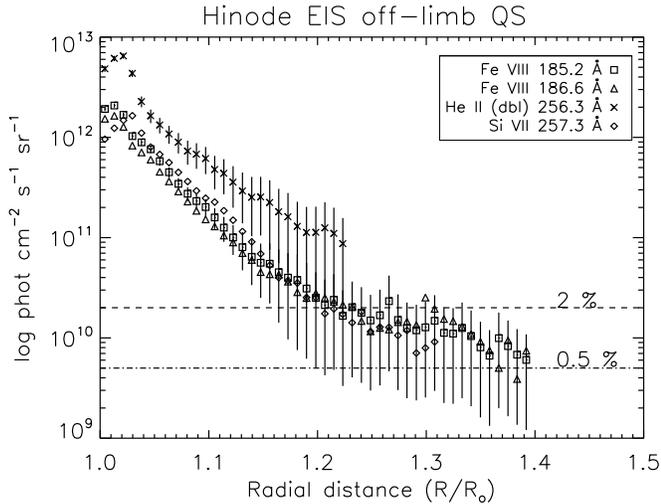}}
\caption{The observed radial fall-off in the cooler EIS lines 
(cf. bottom of Figure~\ref{fig:eis_ol}). The two dashed lines indicate 
2\% and 0.5\% levels from the on-disk values.}
\label{fig:eis_ol2}
\end{figure}
% Figure~\ref{fig:eis_ol2}

Considering the off-limb observations we have presented in this paper, 
as they are rather unique (using the bottom part of the EIS slit), 
there is no available information  on the instrument performance.
However, we can also assess for possible levels of stray light by measuring the 
off-limb behavior in the cooler lines, where negligible signal from the outer corona
would normally be expected.  The strongest line is at 256.3~\AA,
a complex blend of \ion{He}{ii} and several coronal lines
\citep[see section 3.3. in ][]{delzanna_etal:2011_flare}.
We could only deblend the  \ion{Si}{x} contribution using a  branching ratio 
with the 261~\AA\ line,  but the signal was such this could only be feasible up to 
1.2~\rsun. We therefore consider this line not suitable for off-limb measurements of the
stray light.
 We have therefore considered all the other strongest cool transition-region lines,
mostly from \ion{Fe}{viii} and \ion{Si}{vii}, recorded for this observation. 
There is some signal in these lines at various distances up to 1.4~\rsun, as shown in 
Fig.~\ref{fig:eis_ol2}. The 2\% level from the on-disk intensities is shown with the 
upper dashed curve. It is clear that in this case a 2\% level of stray light at 1.5~\rsun\ 
is an overestimate.
Considering that some of the measured signal could actually be real, 
a more conservative $5 \times 10^{-3}$ (lower dashed curve) should be used. 

In conclusions, on the basis of off-limb AIA and EIS observations, a 
level of stray light between $5 \times 10^{-3}$  and 2\%  at 1.5~\rsun\ is possible.
Regarding the amount of stray light that would be acceptable by COSIE-C,
we note that we would require that 
 the coronal emission at say 3~\rsun\  should be greater than the fluctuations in the
scattered disk emission $SD$. Assuming Poisson noise on the number of detected 
photons $D$, we therefore require that $D(3~\rsun) >  \sqrt{SD(3~\rsun)}$.
On the basis of Figure~\ref{fig:cosie_qs} and Eq.~1, 
at around 3~\rsun\  we expect to detect about 2 photons/pixel/s, i.e. 
120 photons/pixel for a 60s exposure time. The disk signal is about 10$^5$  photons/pixel/s,
so we would require the stray light to be less than $3 \times 10^{-3}$ the disk
radiation, at  3~\rsun. This appears achievable, considering the above 
discussion. We would in fact expect the stray light to decrease with 
radial distance from the Sun, and if we assume that this decrease follows the
observed decrease in the  MXUVI signal, a 1\% level at 1.5~\rsun\  would already
provide the required stray light at 2~\rsun. A 2\% stray light at  1.5~\rsun\
would be $6 \times 10^{-3}$ at 2~\rsun\ and most likely less than the requirement 
at  3~\rsun.
If stray light  will turn out to be higher than $3 \times 10^{-3}$ at  3~\rsun,
 one would have to  reduce the spatial resolution (i.e. binning) and/or increase  
the exposure time (60s) to detect the coronal signal in quiescent streamers.

\section{Some estimates for other regions}

Aside from quiet Sun streamers, perhaps the most important 
science targets for COSIE-C will be active region streamers.
To estimate what kind of signal might be expected, we 
 have extended the search on  UVCS observations during the first year of operations,
to try and find streamers above active regions. We found only
one useful observations on 1996 June 1, also shown in Figure~\ref{fig:composite}. 
This is because the other active region observations had very little 
signal, either because of too short exposure times and/or narrow slits, or 
only observed below 2~\rsun. The selected streamer was above two active regions,
a compact small one and a diffuse one. The streamer is not as bright as 
streamers observed later on when the solar activity increased, but provides an 
idea of how the signal increases above an AR.
The radiances of the lines are shown in Figures~\ref{fig:mg_10},\ref{fig:si_12},\ref{fig:lyb}.
The increases are significant,  
especially those of the hotter \ion{Si}{xii} 521~\AA\ line, which is brighter
by over  an order of magnitude closer to the limb, as one would expect. 
However, the radiances  at 3.1~\rsun\  are less than a factor of two higher
than those of the quiet Sun streamers. We therefore expect 
similar increases in the COSIE-C signal, compared to the quiet Sun estimates.
With high solar activity, the  COSIE-C signal would increase much more. 
 
As we have mentioned, CMEs are also a primary target for COSIE-C.
Many CMEs will also be bright in the COSIE passband.
We are developing  models to study the COSIE signal, but a full 
discussion is beyond the scope of the present paper, as it is quite a complex issue.
 
What we can say, on the basis of previous observations, is that 
the fluxes will vary enormously from event to event, and different structures
(cooler and hotter) will be visible by COSIE.
For example, UVCS spectra of CME cores generally show plasma at
transition region temperatures, and COSIE images are likely to be dominated by
the O V 193 \AA\/ multiplet \citep[see, e.g.][]{ciaravella_etal:2000}.
Higher temperature emission is seen in AIA 195, 211
and 335~\AA\ images \citep[cf. ][]{ma_etal:2011}  and UVCS spectra
show  CME-driven shock waves \citep{raymond_etal:2000}.
Flare-CME current sheets are seen in high temperature lines
such as [Fe XVIII] and Fe XXIV  \citep[see, e.g.][]{bemporad_etal:2006,warren_etal:2018}, 
and the leading edges of CMEs
appear in several AIA bands \citep[see, e.g.][]{kozarev_etal:2011}
and in coronal emission lines observed with UVCS, so they should be easily
detectable with COSIE's higher sensitivity.
For example, \cite{giordano_etal:2013} presented an UVCS catalog of CME events
from 1996 to 2005, where a number of coronal lines as the 
Si XII doublet were clearly observed. We can therefore predict that COSIE
would easily observe such events.

Finally, we note that coronal holes are not a primary scientific target for the COSIE mission.
Their densities in the outer corona are significantly lower than 
those of the quiet Sun streamers.
For example, \cite{abbo_etal:2010} measured  a density of 1.1$\times$10$^6$ cm$^{-3}$ at
2.2~\rsun\ in a streamer, while they found a value of 
6.5$\times$10$^5$ cm$^{-3}$  in a region that would generally be considered part
of the coronal hole  (their region 4, though it might be a projection of another streamer).
Also, coronal holes have lower  temperatures, so the EUV  
signal would be significantly lower than that for the QS streamers,
probably by a factor of 10--100, in reasonable agreement with the model of 
\cite{shen_etal:2017}.
If stray light were not a factor, increasing exposure times and spatial binning 
would make coronal holes observable with COSIE-C.

\section{Conclusions}

An EUV coronagraph such as COSIE has a great potential 
for novel observations of  the outer corona, especially 
between 1.5 and 3~\rsun, where many fundamental processes affecting the 
generation of the solar wind but also the evolution of 
structures in the low corona are taking place.
This region is still nearly  unexplored. 

We have a long heritage of coronagraph observations at  visible wavelengths, which 
has notorious difficulties in reducing the disk light and in 
 observing the corona close to the Sun. 
This is not an issue for  COSIE in its coronagraph mode, which will allow, with 
the use of a filter, both on-disk and off-limb observations out to  3.3~\rsun.

It might seem surprising, but we have shown here that significant 
signal in lines collisionally excited is still present at least until 
3.1~\rsun, i.e.  2~\rsun\ above the photosphere. 
The few SOHO UVCS  observations we were able to find clearly show this.
We are confident about these results. 
We have also clearly shown that the same EUV lines of the COSIE
instrument  have been easily observed by Hinode EIS in a relatively 
quiet period in 2007 out to 1.5~\rsun.

The main conclusion of the present study is that, on the basis of current 
baseline designs, COSIE-C exposures of the order of tens of  seconds will be sufficient
to observe  quiescent streamers at 3~\rsun,
 if stray light issues will turn to be negligible.
 This is unprecedented (except for images during eclipses), when 
considered in combination with a spatial resolution of 3\arcsec.

 We have briefly discussed stray light issues. As there are discrepancies 
in assessing the stray light in current instruments, and stray light 
very much depends on the actual optical layout and components (e.g. filters), we defer
an assessment to a future study, noting that our current knowledge on the basis 
of AIA and EIS performances indicates that COSIE-C, with improved 
filters and micro-roughness of the main optical surface,
should have a negligible stray light.

A secondary main conclusion is that variations in the outer corona of the 
electron densities and temperatures obtained from models 
fail to explain the observations. 
It is surprising, but the ionization temperature of the outer corona 
seems to be relatively constant out to 3.1~\rsun, at least on the 
basis of the few observations from UVCS and EIS presented here. That the temperature
is nearly isothermal and constant with height was known from previous 
SOHO CDS and Hinode EIS measurements, but only up to 1.2~\rsun. 

The relative radiances between the coronal lines and the 
\ion{H}{i}  Lyman $\beta$  observed by UVCS indicate photospheric 
abundances in the quiescent 1996 streamers, using the present model.
The  Lyman $\beta$ line is mostly collisional at 1.4~\rsun, so this
result is largely independent on the photoexcitation and the density distribution of the plasma.
It is interesting to note that this result is in agreement with a 
recent revision by \cite{delzanna_deluca:2017}
 of SOHO SUMER observations near the solar limb during the same period.

There are however several complexities related to the interpretation 
and modelling of the UVCS Lyman and \ion{O}{vi} lines that are significantly photo-excited, 
and which could also affect these  elemental abundance measurements 
at larger distances, so these abundance measurements will need to be confirmed.

\acknowledgements
GDZ  acknowledges support from  STFC (UK) and from SAO 
during his visits  to CfA.
We thank the AIA team, in particular Wei Liu, for pointing out
when long exposures were taken in the AIA bands. 
We also thank D. Seaton for providing information about the 
SUVI effective area. 
Finally, we thank the anonymous referee for the constructive 
comments, which helped us to improve the paper.

\bibliographystyle{apj}
\bibliography{../../bib.bib}

\end{document}